%%%%%%%%%%%%%%%%%%%%%%%%%%%%%%%%%%%%%%%%
\newif\ifbbB\bbBfalse           				     %%%
%%%       BLACKBOARD BOLD FONT          %%%
%%%     Comment-out the next line if you      %%%
%%%      do NOT have mssb fonts:              %%%
 \bbBtrue                             %%%
%%%%%%%%%%%%%%%%%%%%%%%%%%%%%%%%%%%%%%%%

\input harvmac
\input tables
% \draftmode
% \input psfig
%\newcount\figno
%\figno=0
%\def\fig#1#2#3{
%\par\begingroup\parindent=0pt\leftskip=1cm\rightskip=1cm\parindent=0pt
%%\baselineskip=11pt
%\global\advance\figno by 1
%\midinsert
%\epsfxsize=#3
%\centerline{\epsfbox{#2}}
%\vskip 12pt
%{\bf Fig. \the\figno:} #1\par
%\endinsert\endgroup\par
%}
%\def\figlabel#1{\xdef#1{\the\figno}}
%\def\encadremath#1{\vbox{\hrule\hbox{\vrule\kern8pt\vbox{\kern8pt \hbox{$\displaystyle #1$}\kern8pt} \kern8pt\vrule}\hrule}}
\def\underarrow#1{\vbox{\ialign{##\crcr$\hfil\displaystyle
 {#1}\hfil$\crcr\noalign{\kern1pt\nointerlineskip}$\longrightarrow$\crcr}}}

\overfullrule=0pt

\ifbbB
  \message{If you do not have msbm blackboard bold fonts,}
  \message{change the option at the top of the tex file.}
  \font\blackboard=msbm10 % scaled \magstep1
  \font\blackboards=msbm7 \font\blackboardss=msbm5
\newfam\black \textfont\black=\blackboard
\scriptfont\black=\blackboards \scriptscriptfont\black=\blackboardss
\def\Bbb#1{{\fam\black\relax#1}}
\else
\def\Bbb{\bf}
\fi

\def\bar{\overline}
\def\I1{\relax\ifmmode\hbox{1\kern-.4em 1}\else{1\kern-.4em 1}\fi}

\def\bJ{{\Bbb J}}

\def\b1{{\Bbb 1}}
\def\bZ{{\Bbb Z}}
\def\USp{{U\!Sp}}

\def\o{\over}
\def\Tr{ \hbox{\rm Tr}}

\def\bra{\langle}
\def\ket{\rangle}

% \vskip 0.5in\centerline{\titlefont #2}\abstractfont\vskip .5in\pageno=0}
%%%%%%%%%%%%%%%%%%%%%%%%%%%%%%%%%%%%%
\def\np#1#2#3{{\it Nucl. Phys.} {\bf B#1} (#2) #3}

\def\plb#1#2#3{{\it Phys. Lett.} {\bf #1B} (#2) #3}
\def\prl#1#2#3{{\it Phys. Rev. Lett.} {\bf #1} (#2) #3}

\def\jhep#1#2#3{{\it J. High Energy Phys.} {\bf #1} (#2) #3}

\def\cmp#1#2#3{{\it Comm. Math. Phys.} {\bf #1} (#2) #3}

\lref\swone{N. Seiberg and E. Witten, ``Monopole condensation and
confinement in N=2 supersymmetric Yang-Mills theory,'' hep-th/9407087,
\np{426}{1994}{19}.}

\lref\ckm{G. Carlino, K. Konishi and H. Murayama, ``Dynamics of
supersymmetric $SU(N_c)$ and $\USp(2n_c)$ gauge theories,''
hep-th/0001036, \jhep{\bf 0002}{2000}{002}; ``Dynamical symmetry
breaking in supersymmetric $SU(N_c)$ and $\USp(2n_c)$ gauge theories,"
hep-th/0005076, \np{590}{2000}{37}.  }

\lref\isone{K. Intriligator and N. Seiberg, ``Phases of $N=1$
supersymmetric gauge theories in four dimensions,'' hep-th/9408155,
\np{431}{1994}{551}.}

\lref\istwo{K. Intriligator and N. Seiberg, ``Duality, monopoles,
dyons, confinement and oblique confinement in supersymmetric $SO(N_c)$
gauge theories,'' hep-th/9503179, \np{444}{1995}{125}; ``Lectures on
supersymmetric gauge theories and electric-magnetic duality,''
hep-th/9509066, {\it Nucl.  Phys.  Proc.  Suppl.} {\bf 45BC} (1996)
1.}

\lref\isthree{K. Intriligator and N. Seiberg, ``Phases of $N=1$
supersymmetric gauge theories and electric-magnetic triality,''
hep-th/9506084.}

\lref\arplsei{P. C. Argyres, M. Ronen Plesser and N. Seiberg, ``The
moduli space of vacua of $N=2$ SUSY QCD and duality in $N=1$ SUSY
QCD,'' hep-th/9603042, \np{471}{1996}{159}.}

\lref\arplsha{P. C. Argyres, M. Ronen Plesser and A. D.
Shapere,``$N=2$ moduli spaces and $N=1$ dualities for $SO(N_c)$ and
$\USp(2N_c)$ Super QCD,'' hep-th/9608129, \np{483}{1997}{172}.}

\lref\swtwo{N. Seiberg and E. Witten, ``Monopoles, duality and chiral
symmetry breaking in N=2 supersymmetric QCD,'' hep-th/9408099,
\np{431}{1994}{484}.}

\lref\seiberg{N. Seiberg, ``Electric-magnetic duality in
supersymmetric nonabelian gauge theories,'' 
hep-th/9411149, \np{435}{1995}{129}.  }

\lref\argdoug{P. C. Argyres and M. R. Douglas, `` New Phenomena in
$SU(3)$ supersymmetric gauge theory,'' hep-th/9505062,
\np{448}{1995}{93}.}

\lref\scftone{P. C. Argyres, M. R. Plesser, N. Seiberg and E. Witten,
`` New $N=2$ superconformal field theories in four-dimensions,''
hep-th/9511154, \np{461}{1996}{71}.}

\lref\argshap{P. C. Argyres and A. D. Shapere, `` The vacuum structure
of $N=2$ superQCD with classical gauge groups,'' hep-th/9509175,
\np{461}{1996}{437}.}

\lref\brandland{A. Brandhuber and K. Landsteiner, `` On the
monodromies of supersymmetric Yang-Mills theory with gauge group
$SO(2n)$,'' hep-th/9507008, \np{358}{1995}{73}.}

\lref\dougshenk{M. R. Douglas and S. H. Shenker, ``Dynamics of $SU(N)$
supersymmetric gauge theory,'' hep-th/9503163, \np{447}{1995}{271}.}

\lref\eguhorione{T. Eguchi, K. Hori, K. Ito and S.-K. Yang, ``Study of
$N=2$  superconformal field theories in four dimensions,''
hep-th/9603002, \np{471}{1996}{430}.}

\lref\eguhoritwo{T. Eguchi and K. Hori, ``$N=2$ superconformal field
theories in four-dimensions and A-D-E classification,''
hep-th/9607125.}

\lref\ktcp{ K. Konishi and H. Terao, ``CP, charge fractionalizations
and low-energy effective actions in the $SU(2) $ Seiberg-Witten
theories with quarks," hep-th/9707005, \np{511} {1998} {264}.}

\lref\GKT{G. Carlino, K. Konishi and H. Terao, ``Quark number
fractionalization in supersymmetric $SU(2)\times U(1)^{N_f} $ gauge
theories," hep-th/9801027, \jhep {04} {1998} {003}.}

\lref\argfar{P. C. Argyres and A. E. Faraggi, ``Vacuum structure and
spectrum of $N=2$ supersymmetric $SU(N)$ gauge theory,''
hep-th/9411057, \prl{74}{1995}{3931}.}

\lref\klemm{ A. Klemm, W. Lerche, S. Theisen and S. Yankielowicz,
``Simple singularities and $N=2$ supersymmetric Yang-Mills theory,''
hep-th/9411048, \plb{344}{1995}{169}.}

\lref\argpltwo{P. C. Argyres, M. R. Plesser and A. D. Shapere,
``The Coulomb phase of $N=2$ supersymmetric QCD,'' hep-th/9505100,
\prl{75}{1995}{1699}.} 

\lref\hananyoz{A. Hanany and Y. Oz, `` On the quantum moduli space of N=2
supersymmetric gauge theories," hep-th/9509176, \np{466}{1996}{85}.}

% \lref\callias{C. Callias,``Axial Anomalies and Index Theorems on Open
% Spaces," \cmp {62} {1978} {213}, E. Weinberg, ``Fundamental Monopoles 
% in Theories with Arbitrary Symmetry Breaking," {\it Nucl. 
% Phys.}{\bf B203} (1982) {445}, J. de Boer, K. Hori and Y. Oz,
% ``Dynamics of N=2 Supersymmetric Gauge Theories in Three Dimensions,"
% hep-th/9703100, {\it Nucl.  Phys.} {\bf B500} (1997) {163}.}

\Title{\vbox{\baselineskip12pt
%\hbox{hep-th/0104064}
\hbox{SWAT/288}\hbox{IFUP-TH 14/2001}\hbox{UCB-PTH-01/13}\hbox{LBNL-47700}}}
{\vbox
{\centerline {Vacuum Structure and Flavor Symmetry Breaking}
\bigskip
\centerline {in Supersymmetric $SO(n_c)$ Gauge Theories}} } 
\smallskip
\centerline{Giuseppe Carlino $^{a}$, Kenichi Konishi $^{b,c}$, S. Prem Kumar $^{a}$}
\centerline{and Hitoshi Murayama$^{d,e}$}
\smallskip
\centerline{\it (a) Dept. of Physics, University of Wales Swansea, Swansea, SA2 8PP, UK}
\centerline{\it (b) Dip. di Fisica, Universit\`a di Pisa, Via Buonarroti, 2, Ed.B, 56127 Pisa, Italy }
\centerline{\it (c) Istituto Nazionale di Fisica Nucleare,  Sezione di Pisa }
%\centerline{\it (d) Department of Physics,  University of Washington, Seattle WA 98195-1560, USA}
\centerline{\it (d)Dept. of Physics, University of California, Berkeley CA 94720, USA}
\centerline{\it (e)Lawrence Berkeley National Laboratory, Cyclotron Road, Berkeley CA94720 USA}
\medskip
\noindent
%\vskip .2in
%\bigskip\bigskip
%\baselineskip 18pt

%: Abstract
\centerline{\bf Abstract}
We determine the vacuum structure  and phases of $N=1$ theories obtained via a mass $\mu$ for the adjoint chiral superfield in $N=2$, $SO(n_c)$ SQCD. 
For large number of flavors these theories have two groups of vacua. 
The first exhibits dynamical breaking of flavor symmetry $\USp(2n_f) \rightarrow U(n_f)$ and arises as a relevant deformation of a non-trivial superconformal theory. These are in the confined phase.
The second group, in an IR-free phase with unbroken flavor symmetry, is produced from a Coulomb branch singularity with Seiberg's dual gauge symmetry.
In the large-$\mu$ regime both groups of vacua are well-described by dual quarks and mesons, and dynamical symmetry breaking in the first group occurs via meson condensation. 
We follow the description of these vacua from weak to strong coupling and demonstrate a nontrivial agreement between the phases and the number of vacua in the two regimes.
We construct the semiclassical monopole flavor multiplets and argue that their multiplicity is consistent  with the number of $N=1$ vacua.

\Date{April 2001}

%: Introduction  and Summary
\baselineskip 18pt
\newsec{Introduction  and Summary}

Confinement and chiral symmetry breaking are the two central features of strong-coupling dynamics in non-Abelian gauge theories in general and QCD in particular.
In their seminal work on the non-perturbative dynamics of $SU(2)$, $N=2$ supersymmetric, pure
gauge theories \swone, Seiberg and Witten explicitly demonstrated that confinement in the corresponding $N=1$ theories could be understood as a dual Higgs mechanism, i.e. the condensation of magnetic monopoles.
In their subsequent work \swtwo\ they also showed that in theories with matter hypermultiplets, flavor symmetries are broken dynamically by the condensation of monopole multiplets transforming under certain representations of the flavor group.
Although their analysis of the low-energy effective action and confinement mechanism was extended to
more general theories (\argshap\ \argfar\ \klemm\ \argpltwo\ \hananyoz), the dynamics of flavor symmetry breaking for more general gauge theories has only recently been investigated in \ckm\ for $SU(n_c)$ and $\USp(2n_c)$ gauge groups.
Similar techniques, when applied to $SO(n_c)$ gauge theories, yield results with some unexpected features. 
It is the purpose of this paper to discuss the interesting aspects of the dynamics of these theories. One of our main objectives is to explore the patterns of flavor-symmetry breaking and identify the dynamical
mechanisms involved.

The models we discuss have $N=1$ supersymmetry and are constructed by perturbing $N=2$ supersymmetric $SO(n_c)$ gauge theory with $n_f$ hypermultiplets in the vector representation.
The $N=1$ preserving perturbation corresponds to a simple mass-deformation via a mass-term for the adjoint chiral $N=1$ superfield $\Phi$ in the $N=2$ vector multiplet. 
The Lagrangean for this theory is given by
\eqn\lagrangian{ {\cal L}= {1\over 8 \pi} {\rm Im} \, \tau_{{\rm cl}} \left[\int d^4 \theta \, \Phi^{\dagger} e^V \Phi +\int d^2 \theta\,{1\over 2} W W\right] + {\cal L}^{({\rm quarks})} + \Delta {\cal L} ,}
where the adjoint mass term
\eqn\Nonepert{\Delta {\cal L}= \int \, d^2 \theta \,\mu \,\Tr \, \Phi^2}
reduces the supersymmetry to $N=1$, and 
\eqn\lagquark{ {\cal L}^{({\rm quarks})} = \sum_i \,  \int d^4 \theta \, \{ Q_i^{\dagger} e^V Q_i + {\tilde
Q_i}^{\dagger} e^{ {\tilde V}} {\tilde Q}_i \} + \int d^2 \theta \, \{\sqrt{2} {\tilde Q}_i \Phi Q^i + m_i {\tilde Q}_i Q^i \} }
describes the interactions of the $n_{f}$ flavors of hypermultiplets (``quarks''). 
The complexified bare coupling constant is $\tau_{{\rm cl}} \equiv {\theta_0 \over\pi} + {8 \pi i \over g_0^2}$.
The pairs $(Q_a^i, {\tilde Q}_a^i)\equiv (Q_a^i, Q_a^{n_f+i})$ $(i=1,2, \ldots, n_f)$ make up the $N=2$ hypermultiplets in the vector representation of the $SO(n_c)$ gauge group.
In the absence of quark masses, the theory also has a global $\USp(2n_f)$ symmetry under which
the pair $(Q_a^i, {\tilde Q}_a^i)$ transforms as a $2n_f$-plet. 
The $N=1$ chiral and gauge superfields $\Phi= \phi \, + \, \sqrt2 \, \theta\,\psi + \, \ldots \, $, and $W_{\alpha} = -i \lambda \, + \, {i \over 2} \, (\sigma^{\mu} \, {\bar\sigma}^{\nu})_{\alpha}^{\beta} \, F_{\mu \nu} \, \theta_{\beta} + \, \ldots$ are both in the adjoint representation of the gauge group.

As in \ckm\ we shall consider adding, besides the adjoint mass, small {\it generic} nonvanishing bare masses $m_i$ for the hypermultiplets (``quarks'').
Non-zero quark masses in the $N=2$ theory lift the flat directions associated with Higgs/mixed branches, leaving a Coulomb branch with isolated singularities.
As usual, a subset of these singular points are special in that they yield $N=1$ supersymmetric vacua upon introducing the adjoint mass.
Thus we obtain a finite number of isolated $N=1$ vacua -- keeping track of this number in various
regimes of the $(\mu,m_i)$ parameter space allows us to perform highly non trivial checks of our analysis.

For small adjoint masses $\mu<<\Lambda_{N=2}$ and $m_i \to 0$ ($\Lambda_{N=2}$ is the dynamical scale of the $N=2$ theory) we find that the $N=1$ vacua are produced as perturbations of {\it two} singular points on the $N=2$ Coulomb branch:

i) One where the hyperelliptic curve exhibits critical behavior of the type $y^2 \propto x^{n_f+4}$ for $n_f$  even and $n_c$ even; $y^2 \propto x^{n_f+2}$ for $n_f$ even and $n_c$ odd; $y^2 \propto    x^{n_f+3}$ for $n_f$  odd and $n_c$ even or odd.
The light degrees of freedom are mutually non-local and the theory flows to an interacting $N=2$ superconformal theory.
We shall refer to this point as the ``Chebyshev point'' because its position in the $N=2$ moduli space is given by the roots of a Chebyshev polynomial. 

ii) The other singular point is the so-called ``special point'' which was identified by Argyres, Plesser and Shapere (APS) in \arplsha.
At this Coulomb branch singularity the gauge symmetry is enhanced to $SO(2n_f-n_c+4)$ and Seiberg's dual gauge group \seiberg\ \isone\ \istwo, makes an appearance.

The above two points are distinguished from other generic Coulomb branch singularities in that they correspond to points of maximal degeneration of the Riemann surface associated with the hyperelliptic
curves for the theory.
It should be noted that the Chebyshev point associated with the $N=2$ SCFT was not considered in the work of \arplsha.
However, as we will see subsequently it potentially plays an important role in the appearance of Seiberg's dual theory in the $N=1$ limit.\foot{Similar SCFT's were also discovered in \ckm\ for $SU(n_c)$ and $\USp(2n_c)$ theories.}
By analyzing the curves in the vicinity of these points, we find that there are 
\eqn\numb{ {\cal N}_1= (n_c- n_f -2) \cdot 2^{n_f}} 
vacua with $N=1$ SUSY originating from the Chebyshev singularity upon mass perturbation. 
The special point of APS gives rise to 
\eqn\numtw{{\cal N}_2 = {\cal N} - {\cal N}_1,} 
vacua, ${\cal N}$ being the total number of supersymmetric vacua given by 
\eqn\total{ {\cal N} = \sum_{r=0}^{min\{[n_{c}/2], n_{f}\}} w(n_{c}-2r) {}_{n_f}\!C_{r}  +  {}_{n_f}\!C_{n_c/2}.}
Here $w(N)$ is the Witten index for $SO(N)$ gauge group with $w(N)= N-2$ for $\quad N \ge 5$ and $w(N)=4,\, 2,\, 1,\, 1,\, 1,\,$ for $N= 4,\, 3,\, 2,\, 1,\,0,\,$ respectively.
The last term in Eq. \total\ is present only for $2n_f \ge n_c$ and $n_c= {\rm even}$. 
The physics of these $SO(n_c)$ gauge theories produced by perturbing the two Coulomb branch singularities in the regime where the adjoint mass $\mu \ll \Lambda_{N=2}$, $m_i \to 0$ and $2n_f>n_c-2$, can be summarized as in  the Table~1.  
The interacting CFT on the $N=2$ Coulomb branch flows to vacua in the confining phase upon introducing the relevant perturbation corresponding to the adjoint mass. 
On the other hand, the theory at the special point flows to a free magnetic phase or non-abelian Coulomb phase (depending on the number of flavors) in the limit where the quarks are strictly massless. For finite quark masses, it yields a set of isolated vacua in Higgs, Coulomb and confining phases.
\bigskip
%: Table 1
\begintable
 Label | Degrees of freedom | Eff. gauge group | Phase | Flavor group \cr
 First group | Mutually nonlocal | -- | Confining | $U(n_f)$  \crnorule
 | | | (deformed SCFT) | \cr
 Second Group | Dual quarks | $SO(2n_f-n_c+4)$ | Free magnetic or | $\USp(2n_f)$  \crnorule
 | | | non-abelian Coulomb |
\endtable~{\bf Table 1:} Phases of the $N=1$ $SO(n_c)$ theory with $2n_f$ flavors and $\mu\ll\Lambda_{N=2}$.
\bigskip
As we discuss below, an extremely non-trivial check of this picture is obtained by analyzing  the theory in a completely different limit, namely $\mu>>\Lambda_{N=2}$. 

One feature that distinguishes the two groups of vacua is that the first group exhibits dynamical flavor symmetry breaking to $U(n_f)$, while the full $\USp(2n_f)$ global symmetry remains unbroken in the second group.

Since we do not have a Lagrangean description of the low energy effective theory at the Chebyshev point, the symmetry breaking pattern can only be obtained
by analyzing the theory at large $\mu>>\Lambda_{N=2}$ where we do have a useful effective description of the theory.
In the large-$\mu$ regime, the adjoint scalar gets frozen out and the theory may be described in terms of mesons and dual quarks \seiberg\ \isone\ \istwo, with a classical superpotential for the mesons obtained by integrating out the adjoint scalar.
The resulting vacuum structure of the large-$\mu$ theory mirrors the small-$\mu$ regime outlined above. In particular we find two groups of vacua -- one which is $\USp(2n_f)$-symmetric, while the other is only $U(n_f)$-symmetric due to the dynamical condensation of mesons.
Moreover we find a total of precisely ${\cal N}_1=(n_c-n_f-2) \cdot 2^{n_f}$ vacua with $U(n_f)$ flavor symmetry, and this allows us to identify these theories as the large-$\mu$ counterparts of the Chebyshev vacua that we encountered above. 

The second group of vacua, with unbroken flavor symmetry, is in the non-Abelian free-magnetic phase.  In the large-$\mu$ regime the low energy degrees of freedom are the dual quarks and mesons, whose
interactions are described by an infrared-free $SO({\tilde n}_c) = SO(2n_f-n_c+4)$ theory. There are no meson condensates and thus no dynamical symmetry breaking takes place.
In the small-$\mu$ regime these are described by a local effective Lagrangean which was identified by APS in \arplsha.
The multiplicity of the second group of vacua also matches that found from the analysis of the curve at the special point.

Vacuum counting in both the large and small-$\mu$ regimes thus provides a non-trivial demonstration of the fact that the  Chebyshev point does indeed yield $N=1$ supersymmetric vacua which were not considered in \arplsha. 
Importantly, the behavior of the meson VEVs in the decoupling limit ($\mu\to \infty$ with the $N=1$ dynamical scale fixed) is rather interesting. The meson-condensates vanish and the two
classes of vacua merge. 
This suggests that the physics of the Chebyshev point on the $N=2$ Coulomb branch needs to be understood in order to explain the origin of Seiberg's dual degrees of freedom (the mesons in particular) in $N=1$ SUSY-QCD. 

Though we understand the pattern of flavor symmetry breaking at the Chebyshev point, we do not have a clear picture of the microscopic mechanism involved. In particular, while we know that in the large $\mu$ regime the relevant mechanism is the dynamical condensation of mesons, we have no clear understanding of the light degrees of freedom that condense in the small $\mu$ description of the theory. However, in the case  where the quark masses are nonvanishing and equal, i.e. $m_i = m_0 \neq 0$, we can accurately analyze the low energy dynamics in the vicinity of the Chebyshev point. Perhaps this will shed some light on the physics in the $m=0$ case as well.
The flavor symmetry group of the underlying theory is now broken explicitly to $U(n_f)$. Analysis of the theory at large $\mu$ reveals that the first group of vacua splits    
into several subgroups each labelled by an integer, $r=0,1,2,\ldots, [n_f/2]$, and with flavor symmetry $U(r)\times U(n_f-r)$.

In the small-$\mu$ regime on the other hand, the form of the hyperelliptic curve at these singular points confirms this picture. In particular, the curve becomes critical at each of these points and the criticality of the curves (following the classification of \eguhorione\ and \eguhoritwo) suggests that these theories are in the same universality class as the IR-free theories encountered in the work of Argyres, Plesser and Seiberg \arplsei\ at the roots of the $r-$Higgs branches of $SU(n_c)$ SQCD.
Each of these theories is described by a {\it local}  effective gauge theory \`a la Argyres-Plesser-Seiberg, with gauge group $SU(r) \times U(1)^{[{n_c\over 2}] -r+1}   $   and $n_f$ (dual)  quarks in the fundamental representation of $SU(r)$. Indeed, the gauge invariant composite VEVS  characterizing these theories differ by  some powers of $m$, and the validity of each effective theory is limited  by  small fluctuations  of order of $m$ around each vacuum. 
In the limit $m\to 0$   these points  in the quantum moduli space (QMS) with different symmetry properties collapse into one single point. In this limit, at the Chebyshev point the criticality of the curve is of the form $y^2\propto x^{n_f+4}$  for $n_c$ even and $y^2\propto x^{n_f+3}$ for $n_c$ odd, indicating the appearance of an interacting SCFT (\eguhorione\ \eguhoritwo).

The phases and degrees of freedom in the first group of vacua with
$m_i=m_0\neq 0$ and $\mu<<\Lambda_{N=2}$ are summarized in Table 2. 
\bigskip
%: Table 2
\begintable
 Label $(r)$ | Degrees of freedom | Eff. gauge group | Phase | Flavor group \cr
 $r=0$ | monopoles | $U(1)^{[{n_c\over 2}]}$ | confining | $U(n_f)$ \cr 
 $r=1$ | monopoles | $U(1)^{[{n_c\over 2}]}$ | confining | $U(n_f-1) \times U(1)$   \cr
 $r=2, \ldots, [n_{f}/2] $ | dual quarks | $SU(r) \times U(1)^{[{n_{c} \over 2}] - r+1}$ | confining | $U(n_{f}-r) \times U(r)$ 
\endtable~{\bf Table 2} The first group of vacua of $SO(n_c)$ gauge theory with  $2n_f$ flavors and $m_{i}=m_0 \ne 0$.
\bigskip

One of the most interesting outcomes of the large $\mu$ analysis discussed above, is the identification and tracking -- at large and small $m_i$'s -- of vacua in distinct phases, such as Higgs, confinement (magnetic Higgs) or Coulomb.  
In contrast to the theories considered in \ckm, $SO(n_c)$ gauge theories with quarks in the vector representation present a clear distinction between Higgs and confinement phases since the behavior of the Wilson loop in the spinor representation is qualitatively different in the two cases. Therefore, unlike other examples (as in \swtwo\ for $SU(2)$) a Higgsed-vacuum in the semiclassical regime (large $m_i$) must remain in the Higgs phase in the strong-coupling regime ($m_i\to 0$) as well. 
Failure to do so would imply a phase transition which in turn is forbidden by holomorphy in $m_i$ or $N=1$ supersymmetry. 
We find nontrivial agreement between phases and vacuum counting in both the semiclassical (large $m_i$) and quantum (small $m_i$) regimes and find results consistent with the absence of any phase transitions. In particular, vacua which appear to be in the Higgs (or Coulomb) phase semiclassically, can be explicitly shown in the strong coupling regime to be in a magnetically confined (or magnetic Coulomb) phase.

The paper will be organized as follows.  We first establish, in Section 2, the number of supersymmetric vacua by analyzing the theory semiclassically.
By $N=1$ supersymmetry and holomorphy in $\mu$ and $m_i$, these results, valid at large $\mu$ and $m_i$, are also correct at small values of these parameters.

Next, in Section 3 we determine the pattern of dynamical symmetry breaking in each vacuum at large $\mu$ (and small generic $m_i$).
In all cases we reproduce the correct number of vacua starting from the known $N=1$ low-energy effective Lagrangean, adding to it the term arising from integrating out the heavy adjoint field $\Phi$, and minimizing the potential.
The task turns out to be quite nontrivial since in the $SO(n_c)$ theory the form of the superpotential and the effective degrees of freedom vary with $n_c$ and $n_f$ and there are many cases to be studied separately.

Section 4 and 5 are devoted to the study of the theory at small $\mu$ and small $m_i$'s.
The analysis requires the exact solution of these theories in the $N=2$ ($\mu=0$) limit \argshap\ in terms of the corresponding hyperelliptic curves.  
First we re-analyze the low-energy effective Lagrangean at the ``special point'' obtained by Argyres, Plesser and Shapere in \arplsha. 
However, this effective theory yields only the $\USp(2n_f)$-symmetric vacua.  
We then perturb certain superconformal (Chebyshev) points on the exact curve, and show that all the vacua with dynamical symmetry breaking to $U(n_f)$ are indeed related to different classes of interacting SCFT's.

We conclude with a discussion on the semiclassical monopole states in $SO(n_c)$ theories, which display certain qualitative differences from the cases of $SU(n_c)$ and $\USp(2n_c)$ theories.
In the latter case, an order of magnitude agreement was found between the multiplicity of semiclassical monopole states and the number of $N=1$ vacua. This fact led the authors of \ckm\ to relate the light degrees of freedom condensing in these vacua with semiclassical monopole multiplets.
A naive attempt to perform a similar analysis for $SO(n_c)$ gauge group seems to fail because of a mismatch in the counting.
We provide an explicit construction of monopole flavor multiplets and show that this puzzle arises only in the case with strictly massless quarks.
Since one obtains isolated $N=1$ vacua only upon introduction of non-zero quark masses, we claim that the contradiction is resolved.
Furthermore, this construction will perhaps provide a clue as to which monopole multiplets condense at the Chebyshev points to dynamically break the flavor symmetry group.

%: Semiclassical Vacua of $SO(n_c)$ SUSY-QCD
\newsec{Semiclassical Vacua of $SO(n_c)$ SUSY-QCD}

We begin by exploring the classical vacuum structure of {\it softly broken} $N=2$ SUSY-QCD (with $N=1$ supersymmetry) with $n_f$ quark hypermultiplets and $SO(n_c)$ gauge group.
Although in general the classical picture will be altered by large quantum corrections, in certain situations (when the quark masses $m_i$ are non-zero and large compared to the strong-coupling scale of the theory) the classical analysis may be used to obtain information that is expected to be valid in the full quantum theory as well.
In particular, useful and reliable information on the number of vacua and the symmetry breaking patterns (in some cases) can in fact be obtained from a purely semiclassical analysis. 
As noted earlier, the superpotential for this theory is given by
\eqn\superpson{W= {1 \over 2} \mu \Tr \Phi^2 + \sqrt2 Q_a^i \Phi_{ab}
Q_b^j {\bJ}_{ij} + {1\over 2} m_{ij} Q_a^i Q_a^j,}
where
\eqn\jsymp{{\bJ} = \pmatrix{ {\bf 0} & {\I1} \cr {-\I1} & {\bf 0} } \;
, \qquad \qquad
 m= \pmatrix{ {\bf 0} & {\I1} \cr {\I1} & {\bf 0} } \otimes
{\rm diag} (m_1, m_2, \ldots, m_{n_f}) \; .}

Note that pairs of quark multiplets $(Q_a^i, Q_a^{n_f+i})$ $(i=1,2, \ldots, n_f)$ constitute an $N=2$ hypermultiplet.
The scalar field $\Phi$ belonging to the $N=2$ vector multiplet is in the adjoint representation of the gauge group so that $\Phi_{ab}= t^A_{ab}\Phi^A$,
with
\eqn\songen{ t^A_{ab}={i\over 2}\left[\delta_{a,A}\delta_{b,A+1}-\delta_{a,A+1}\delta_{b,A} \right]}
representing the generators of $SO(n _c)$ rotations in the ${ab}$ plane.

The $\mu=0$ theory (with $N=2$ SUSY) has a $\bZ_{2n_c-2n_f-4} \times SU(2)_R$ $R$-symmetry \foot {The $ U(1)_R$ symmetry of the $\mu =0$ theory is broken by instantons to $\bZ_{4n_c-4n_f-8}$, a $\bZ_2$ subgroup of which is isomorphic to $(-1)^F$ in the Lorentz group.} which is spontaneously broken to $\bZ_2\times SU(2)_R$ by the VEV of $\Tr\phi^2$.
In the $N=1$ theory with $\mu\neq 0$ on the other hand, the adjoint mass explicitly breaks the $R$-symmetry down to $\bZ_2$. 
The additional $\bZ_{n_c-n_f-2}$ discrete symmetry of the parent $N=2$ theory then acts on the vacua of the $N=1$ theory via permutations.

In the $m_i\rightarrow 0$ limit this theory has a $\USp(2n_f)$ flavor symmetry under which the $N=1$ quark multiplets transform as vectors, while all other fields are singlets\foot{The superpotential itself is
in fact invariant under $Sp(2n_f)$ transformations in the massless limit.
The invariance of the kinetic terms however requires the transformations to be unitary as well so that the massless theory has only a $\USp(2n_f)$ global symmetry.}.
Although the $\USp(2n_f)$ flavor symmetry is broken to $U(1)^{n_f}$ for generic, non-zero values of
the quark masses, when the quark masses are all taken to be equal and nonvanishing, $m_i=m_0\neq 0$, a $U(n_f)$ global symmetry is preserved.

As we will see the global $\USp(2n_f)$ symmetry of the $m_i=0$ theory and the $U(n_f)$ symmetry of the $m_i=m_0\neq 0$ theory, will also be broken spontaneously (or dynamically) at various $N=1$ vacua where certain fields (corresponding to the light degrees of freedom at the Coulomb branch singularities of the $\mu=0$ theory) condense and obtain VEVs.

The classical vacuum structure can be obtained by solving the D and
F-term equations:
\eqn\dphi{[\Phi,\Phi^{\dagger}]=0,} 
\eqn\dsq{{\rm Im}Q_a^{i\dagger}Q^b_i=0,} 
\eqn\fphi{\sqrt 2 Q_a^{i\dagger}Q_b^i\bJ_{ij}+2\mu\Phi^{ab}=0,}
\eqn\fsq{\sqrt 2 \Phi^{ab}Q_b^j\bJ_{ij}+m_{ij}Q_a^j=0.}
We will assume that the quark masses take on generic {\it non-zero}
values.

Since the D-term equation \dphi\ requires $\Phi$ to live in the Cartan subalgebra of the gauge group, we may use general gauge rotations to write $\Phi$ in the form
\eqn\vevphi{\Phi= \pmatrix{ \pmatrix{ &\phi_1 \cr -\phi_1 &} & & & \cr
& \pmatrix{ &\phi_2 \cr -\phi_2 &} & & \cr & & \ddots & \cr & & &
\pmatrix{ &\phi_{[n_c/2]} \cr -\phi_{[n_c/2]} &}}.}
In addition, when $n_c$ is odd, there is a null row and null column in the $\Phi$ matrix.
The F-term condition \fsq\ also implies that non-zero squark fields $Q^i$, $(i=1,2,\ldots,n_f)$ must be
eigenvectors of $\Phi$ with eigenvalues $ m_i/\sqrt 2$, while the squarks $Q^{i+n_f}$, $(i=1,2,\ldots,n_f)$must be eigenvectors with eigenvalue $-m_i/\sqrt 2$.
Since the eigenvalues of $\Phi$ are in
fact $\pm i\phi_i$, the non-zero $\{\phi_i\}$ should be taken to equal
$\{\pm im_i/\sqrt 2\}$, modulo permutations which represent the action
of the Weyl group on a given solution and are therefore gauge
equivalent.  Solutions with different choices of signs for the
$\{\phi_i\}$ can also be shown to be gauge equivalent.  (However, this
is not always true as will be discussed in more detail below.) 
Finally, for every vanishing eigenvalue $\phi_k$ both the eigenvectors
$Q^k$ and $Q^{k+n_f}$ must vanish.  Hence one may classify the
semiclassical vacua of the theory according to the number of non-zero
$\phi_i$'s, or equivalently the number of nontrivial eigenvector pairs
$(Q^i,Q^{i+n_f})$.  The solution for $\Phi$ with eigenvalues
$m_1,m_2,\ldots,m_r,$ is then:
\eqn\vacsona{
\Phi= {1 \over \sqrt 2} \pmatrix{ \pmatrix{ & im_1 \cr 
- im_1 & } &  & & &\cr
    & \pmatrix{ &im_2 \cr -im_2 &} &  & \cr  & & \ddots  && \cr
     &  & & \pmatrix{ &im_r \cr -im_r & \cr} & \cr
     &&&& {\bf 0}}.}
The corresponding squark VEVs for the flavors $i=1,\ldots n_f$ are
given by
\eqn\vacsonb{\eqalign{ 
&\longrightarrow i=1,2\ldots n_f \cr
Q^i_a= &\pmatrix{d_1 & 0 & 0 &\cdots &0 &0\ldots\cr
-id_1 & 0 & 0 & \cdots & 0 &0\ldots\cr
0 & d_2 &0 &\cdots & 0 &0\ldots\cr
0 & -id_2& 0 &\cdots & 0 &0\ldots\cr
\vdots& \vdots & \vdots &\vdots &\vdots&0\ldots\cr
0 & 0 & 0 &\cdots & d_r&0\ldots\cr
0 & 0 & 0 & \cdots & -id_r & 0\ldots\cr
\vdots&\vdots&\vdots & \vdots &\vdots&\vdots\cr}}}
This form of the VEVs is completely constrained (up to an overall
phase) by the eigenvalue equations obtained as a consequence of the
F-term conditions.  The phases may always be set to zero by
independent $SO(2)$ gauge rotations generated by the Cartan subalgebra
leaving the VEV of $\Phi$ invariant.  The fields $Q^{i+n_f}_a$ also
obtain similar VEVs, their magnitudes and phases being constrained by
\dsq\ and \fphi, so that:
\eqn\vacsonc{ Q^{i+n_f}_a= \pmatrix{\tilde d_1 & 0 & 0 &\cdots &0 
&0\ldots\cr
i\tilde d_1 & 0 & 0 & \cdots & 0 &0\ldots\cr
0 & \tilde d_2 &0 &\cdots & 0 &0\ldots\cr
0 & i\tilde d_2& 0 &\cdots & 0 &0\ldots\cr
\vdots& \vdots & \vdots &\vdots &\vdots&0\ldots\cr
0 & 0& 0 &\cdots & \tilde d_r&0\ldots\cr
0 & 0& 0 &\cdots & i\tilde d_r& 0\ldots\cr
\vdots&\vdots&\vdots&\vdots&\vdots&\vdots\cr}}
with
\eqn\vacsond{
|d_r|=|\tilde {d}_r|;
\quad\quad\quad{\rm Re}\;({d_r \tilde{d}_r})= - \mu {m_r \over 2}.}

Clearly, from the form of the above solutions a classical vacuum with
$r$ pairs of nonzero eigenvectors, or equivalently $r$ non-vanishing
$\phi_i$'s has an unbroken $SO(n_c-2r)$ gauge symmetry.  Furthermore,
in the presence of the non-zero $N=2$-breaking term $\mu$, it can be
shown that all quark multiplets charged only under the unbroken gauge
group are {\it massive} in such a vacuum.  The low-energy theory is
thus expected to be in the same universality class as pure
supersymmetric glue with $SO(n_c-2r)$ gauge group.  Such vacua with an
effective $SO(n_c-2r)$ gauge symmetry will henceforth be referred to
as ``$r$-vacua''.  The $r$ nonzero eigenvalues may be chosen in
$\pmatrix{n_{f} \cr r}$ distinct ways, each corresponding to a
distinct classical $r$-vacuum.  By standard arguments, each such
classical theory must yield $w(n_c-2r)$ quantum vacua, $w$ being the
Witten index for the pure $SO(n_c-2r)$ SUSY gauge theory.  For $N \ge 5$ the Witten index $w(N)=N-2$ while for $N=0,\;1,\;2,\;3,\;4$ it takes on the values $1,\;1,\;1,\;2$ and $4$ respectively.  This semi-classical counting therefore gives rise to a total of $\cal {N}$ vacua where \eqn\semic{ {\cal N} = \sum_{r=0}^{min\{[n_{c}/2], n_{f}\}} w(n_{c}-2r) \pmatrix{n_{f} \cr r} + \pmatrix{n_{f} \cr
n_c/2},} As explained below the last term must be included only when
$n_f\ge n_c/2$ for $n_c$ even.  This may be understood by first noting
that for $n_f\ge n_c/2$ and $n_c$ even, the baryon operators which
label gauge-inequivalent vacua are non-vanishing only when $r=n_c/2$. 
Secondly we observe that the squark VEVs get interchanged as
$Q^i\leftrightarrow {Q}^{i+n_f}$ under $m_i\rightarrow -m_i$ which
corresponds to flipping the signs of certain eigenvalues of $\Phi$. 
(Recall that the eigenvalues are determined only up to a sign which can
be gauged away when $r<n_c/2$).  When $r=n_c/2$, under an odd number
of $Q_i\leftrightarrow {Q}_{i+n_f}$ flips the non-zero baryon VEVs
change sign, thus yielding a new gauge-inequivalent set of
${}_{n_f}\!C_{n_c/2}$ vacua.  For $r<n_c/2$ all the baryon VEVs are
identically zero and the sign flips do not yield new vacuum
states\foot{Here we give an alternate explanation for the appearance
of the additional ${}_{n_f}\!C_{n_c/2}$ vacua for $n_c$ even and
$n_f\ge n_c/2$.  This additional set does not appear in the case when
$n_c$ is odd.  As noted earlier, the non-zero eigenvalues $\phi_i$ are
determined only up to a sign.  However, whenever there is at least one
zero-eigenvalue (as is always the case for $n_c$ odd) this sign can be
rotated away by an $SO(n_c)$ gauge element of the form ${\rm
diag}(1,1,\ldots,\sigma_3,1,\ldots,-1)$ where $\sigma_3$ is the Pauli
matrix.  For even $n_c$, zero-eigenvalues are possible only when
$r<n_c/2$.  When $r=n_c/2$ (which can happen only when $n_f\ge n_c/2$)
a solution with $\{\phi_i\}=(-m_1,-m_2,+m_3,\ldots,+m_{n_c/2})$ for
example, can be gauge transformed into $\{\phi_i\}=\{+m_i\}$ by the
action of the $SO (n_c)$ gauge element ${\rm
diag}(\sigma_3,\sigma_3,1,\ldots,1)$.  In general (for $r=n_c/2$)
solutions with an odd number of $\phi_i$'s equal to $-m_i$ (rather
than $+m_i$) can be shown to be gauge equivalent, while solutions with
an even number of $\phi_i$'s equal to $-m_i$ form a distinct
equivalence class.  Since the two classes are gauge {\it
inequivalent}, this leads to an additional set of
${}_{n_f}\!C_{n_c/2}$ vacua.}.

Note that for $ 2n_f \le n_c-5 $, the above series can be summed and
gives
\eqn\none { {\cal N} = {\cal N}_1= (n_c- n_f -2) \cdot 2^{n_f} =
(2n_c- 2n_f -4) \cdot 2^{n_f-1}.  }

One expects the above semiclassical enumeration of vacuum states to be
valid for large hypermultiplet masses (when all the matter fields are
very heavy compared to the $N=2$ dynamical scale).  However,
holomorphy properties of the supersymmetric theory
ensure that the above results for the number of vacua continue to hold
as we smoothly dial the masses to smaller values (compared to the
dynamical scale).  A similar statement applies to the flavor symmetry
breaking patterns for the theory with equal and non-zero
hypermultiplet masses $m_i=m_0\neq 0$.  Recall that the theory with
$m_i=m_0\neq 0$ has a $U(n_f)$ flavor symmetry.  The classical
solutions above imply that this flavor symmetry is spontaneously
broken in the $N=1$ vacuum with $SO(n_c-2r)$ gauge symmetry to
$U(r)\times U(n_f-r)$.  The first factor of $U(r)$ can be understood
as the combined action of flavor and global gauge transformations that
leave the VEVs \vacsonb\ and \vacsonc\ unchanged (for equal quark
masses).  The remaining $U(n_f-r)$ is simply the subgroup of $U(n_f)$
that rotates the $n_f-r$ squarks with vanishing VEVs.  
One can easily argue that this classical pattern of symmetry breaking will be reproduced in the full quantum theory as well, so that in each $r$ vacuum the $U(n_f)$ flavor symmetry is indeed broken to $U(r)\times U(n_f-r)$.
Once again holomorphy guarantees the validity of our
conclusions for small masses as well (at least for all $m_0\neq 0$). 
In the following sections we will explicitly see how this is realized
in the full quantum theory.

We remark that the classical analysis cannot be used to draw any
conclusions about flavor symmetry patterns in the limit of vanishing
quark masses.  In the $m_i\rightarrow 0$ limit, while classically one
would expect the flavor symmetry to be completely restored, quantum
mechanically we will find very different results.

%: Dynamical symmetry breaking at large $\mu$
\newsec{Dynamical symmetry breaking at large $\mu$}

In the previous section we obtained the vacuum structure of softly broken $N=2$ SUSY-QCD in the classical (or semiclassical) regime.  
We will now analyze the same theory in a very different limit-- namely,
when the adjoint mass $\mu$ is taken to be much larger than the
dynamical scale of the $N=2$ theory.  In this limit (when $\mu$ is
large, yet finite), one may consistently integrate out the adjoint
scalar $\Phi$ to obtain a low-energy effective $N=1$ superpotential. 
The latter effective theory can be determined precisely using our
knowledge of the low-energy degrees of freedom of $N=1$, $SO(n_c)$
gauge theories which were extensively studied by Intriligator and
Seiberg in \isone\istwo\ and \isthree.  The theories studied in
\isone\istwo\ and \isthree\ can be recovered from the theories we
study, only in the strict decoupling limit which corresponds to
sending $\mu$ to infinity, holding fixed the effective $N=1$
strong-coupling scale.  Depending on the number
of massless flavors in the vector representation the low energy
degrees of freedom are mesons and/or monopoles and in some cases
exotic composites as well; when the number of flavors is sufficiently
large, the low energy dynamics is described by dual (magnetic) gauge
theories with quarks and gauge-singlet mesons.

The analysis of these theories will reveal the flavor symmetry breaking patterns in both the $m_i\to 0$ and $m_i=m_0\neq 0$ cases.  
We emphasize however, that we will always work with the assumption that the masses are strictly non-zero and then observe the behavior of the symmetry-breaking condensates as we take $m_i/\Lambda_{N=2}\ll 1$.  

In what follows, we summarize the results of this section. The reader who wishes to skip the technical details can jump to the next section. Based on the analysis, we find that the vacua can be classified into two sets:

1) A set of ${\cal N}_1$ vacua with vacuum expectation values that remain finite in the above-mentioned small mass limit.
These vacua are in the confined phase and obtain a superpotential contribution via gaugino condensation. 
For $2n_f > n_c  - 2$ these vacua are appropriately described by mesons and dual quarks. 
The superpotential is induced by integrating out the dual degrees of freedom.
These vacua with `large VEVs' will thus exhibit dynamical flavor symmetry breaking since the order parameters (namely, the mesons) will turn out to have condensates proportional to the dynamical scale of the theory. In the massless quark limit, the mesons transform as the rank-two symmetric tensor under the $USp(2n_f)$ flavor group, and their condensates
break $\USp(2n_f) \rightarrow U(n_f)$.

2) A set of ${\cal N}_2$ ground states where VEVs are proportional to various powers of the masses $m_i$ (in the above-mentioned small mass limit) and where one expects the complete $\USp(2n_f)$ flavor symmetry to remain unbroken in the zero mass limit. This set appears only when $2n_f \geq n_c - 4$.
Depending on the number of flavors, the relevant degrees of freedom are either monopoles, dual quarks and mesons, or other exotic composites.
In contrast to the first group of vacua, the second group can be in any of Higgs, Coulomb or confining phases.
These vacua with `small' VEVs will be loosely referred to as `$\USp(2n_f)$ symmetric' even though this is expected to be strictly true only when the quark masses are zero.

The quantum description of these vacua, obtained in the large $\mu$ limit, agrees with semiclassical expectations both in terms of phase structure and of vacuum counting.

\subsec{Generic case with $ 2n_f \le n_c-5 $, $n_c \ge 4$}

When $\Phi$ is sufficiently heavy we may integrate it out to obtain the effective superpotential for the light degrees of freedom which are known to be meson-like excitations.  The leading effect (in a
$1/\mu$-expansion) of integrating out the adjoint scalar is captured by simply solving the classical equation of motion for $\Phi$ from \superpson.
The classical superpotential is then $-\Tr(\bJ M \bJ M)/2\mu+\Tr(mM)/2$ where we have introduced the gauge-invariant meson fields $M^{ij}=Q^i\cdot Q^j$.  However, to this we must add the dynamically generated superpotential induced by gaugino condensation or instanton effects as in \isone, to obtain
\eqn\dynamic{ W= -{1 \over 2 \mu} \Tr (M\bJ M\bJ) + {1\over 2} \Tr (m M) + {L \over ({\rm det} M)^{1/n_c-2 n_f-2}},}
where
\eqn\rthr{ L= { 1 \over 2} (n_c-2n_f -2) \omega_{n_c-2n_f-2}( 16 \Lambda_{n_c, 2n_f}^{3n_c -2n_f-6} )^{1/n_c-2 n_f-2}.}
Here $\omega_n$ denotes the $n$-th root of unity while $\Lambda_{n_c,2n_f}$ is the strong-coupling scale of the $N=1$ theory obtained by decoupling $\Phi$.  It is related to the $N=2$ dynamical
scale $\Lambda_{N=2}$ via $\Lambda_{N=2}^{2(n_c-2-n_f)}\mu^{n_c-2}=\Lambda_{n_c,2n_f}^{3(n_c-2)-2n_f}$.

Differentiating Eq.~\dynamic\ with respect to the meson fields $M_{ij}$ yields the vacuum equations
\eqn\veqfour{ - {1 \over \mu} (\bJ M \bJ )_{ij} + {1\over 2}m_{ij} - { 1\over n_c-2n_f-2 }\, {L \over (\det M)^{1/(n_c-2n_f-2)+1} } ({\rm Cofac} M)_{ij} =0,} 
where $({\rm Cofac} M)_{ij} = (M^{-1})_{ji} ({\rm det} M)$.
Then using the following parametrization for the meson matrix,
\eqn\ac{ M= \pmatrix{A & B \cr B^T & C}, \qquad A=A^T, \quad C=C^T,}
and
\eqn\mtilde{\tilde{m}={\rm diag}(m_1,m_2,\ldots,m_{n_f})}
we are led to a set of equations for the $n_f\times n_f$ submatrices:
\item{i)} $((B^T)^2-CA)/\mu-\tilde{m}B^T\propto \bf 1$;
\item{ii)} $(B^2-AC)/\mu-\tilde{m}B\propto \bf 1$;
\item{iii)} $(B^T C-CB)/\mu={\tilde m}C$;
\item{iv)} $(-AB^T +BA)/\mu={\tilde m}A$.

\noindent It is easy to see that conditions (i) and (ii) together require $\tilde m$ and $B$ to be commuting matrices so that they are simultaneously diagonal.  On the other hand, the symmetry properties of $A$ and $C$ along with equations (iii) and (iv) lead to $A=C=0$.  The solutions to Eq.~\veqfour\ are therefore of the form
\eqn\Sum{ A=C=0, \qquad B= {\rm diag}(\lambda_1, \lambda_2, \ldots,
\lambda_{n_f}).}
The resulting equations of motion for the $\lambda_i$ are,
\eqn\eqlam{ - {1 \over \mu} \lambda_i^2 + {1\over 2}m_i \lambda_i + X=0,} \eqn\eqlamX{ X= -{ L\over{n_c-n_f-2}} \quad
(\prod_{j=1}^{n_f}\lambda_j)^{-2/(n_c-2n_f-2) }.}
These cannot, however, be solved exactly in general.  Nevertheless, we can certainly determine the number of distinct solutions (or equivalently, the number of distinct vacua) and the symmetry breaking
patterns in certain limiting regimes.  The total number of vacua is always independent of the specific values of the mass parameters.

$\hbox{$\bf{m_i=0}$}$:
In the $m_i \to 0$ limit ($m_i\ll \Lambda_{N=2},\mu$), the solution to \eqlam\ is
\eqn\lam{ \lambda_i= \pm\lambda=\pm \sqrt{\mu X},\quad i=1,2,\ldots,
n_f, }
and
\eqn\qqua{ X \propto {e^{2 \pi i k /(n_c-n_f-2)} \Lambda_{N=2}^2 \, \mu} 
\; , \qquad k=1,2,\ldots, n_c-n_f-2.  }
If $r$ is the number of $\lambda_i$'s equal to $+\lambda$, then there
are $(n_c-n_f-2)\;\pmatrix{n_{f} \cr r}$ such distinct solutions and
the total number of $N=1$ vacua is thus,
\eqn\vac{{\cal N}= (n_c-n_f-2) \sum_{r=0}^{n_f}\pmatrix{n_{f} \cr r}=
(n_c-n_f-2) 2^{n_f}}
which agrees precisely with the total number of the semi-classical
vacua \semic, with $2n_f \le n_c-5.$
However, unlike the classical case in the $m_i\rightarrow 0$ limit,
the quantum-corrected effective action of the large-$\mu$ theory
exhibits a spontaneous breakdown of the global $\USp(2n_f)$ symmetry
since some of the meson VEVs remain non-zero in this limit.  This
disagreement does not lead to any contradiction since the classical
analysis was after all argued to be reliable only when the quark
masses were non-zero.  In particular, the vacuum solutions for the
meson fields are of the generic form
\eqn\OK{ M= \pmatrix{{\bf 0} & \pmatrix{\pm\lambda & 0 &\ldots \cr 0 &
\pm\lambda & \ldots \cr \ldots & \ldots & \ddots} \cr
\pmatrix{\pm\lambda & 0 &\ldots \cr 0 & \pm\lambda & \ldots \cr \ldots
& \ldots & \ddots} & {\bf 0} }.}
We will now show that this preserves a $U(n_f)$ subgroup of the flavor
symmetry.

It is easiest to see this in the case where all the $\lambda_i$'s have
the the same phase (or sign) so that $M=\pmatrix{0 & B \cr B & 0}$ and
$B=\lambda {\I1}$.  A general $\USp(2n_f)$ generator $G$ that leaves
such a solution invariant must satisfy $[G, M]=0$.  Parametrizing $G$
as $\pmatrix{g_1 & g_2\cr g_2^* & -g_1^T}$ with the usual constraints
$g_1^\dagger=g_1$ and $g_2^T=g_2$, we find that $g_1$ and $g_2$ must
satisfy $g_1^T=-g_1$ and $g_2^*=g_2$.  In other words, the purely
imaginary, antisymmetric $g_1$ and real, symmetric $g_2$ are the
unbroken generators.  We can rewrite $G$ in a more familiar form after
performing a similarity transformation $G\rightarrow G^\prime={1\over
\sqrt 2}\pmatrix{{\I1}& {\I1}\cr -{\I1}& {\I1}}\;G\;{\sqrt
2}\pmatrix{{\I1}& {\I1}\cr-{\I1}& {\I1}}^{-1}$ which yields
$G^\prime=\pmatrix{g_1+g_2& 0 \cr 0 & g_1-g_2}$.  Given the
constraints on $g_1$ and $g_2$ this is clearly the most general form
for a $U(n_f)$ generator.

It is now easy to see that even when some of the nonvanishing meson
elements have different signs, the unbroken symmetry group is still
$U(n_f)$.  Suppose that one of the nonvanishing elements has a minus
sign: \eqn\minsign { M^{i, n_f+i}= Q_a^i Q_a^{n_f+i} = - \lambda \, .}
Consider a $\USp(2n_f)$ transformation $P$ which acts as
the $SU(2)$ matrix $\pmatrix {0 & 1\cr -1 & 0}$ in the $(i, n_f+i)$
subspace, but has the trivial action on other elements.
This transforms $ M^{i, n_f+i} \to - M^{i, n_f+i}$, and so changes the
sign of the condensate \minsign.  Therefore $U(n_f)$ transformations
generated by $G$ when combined with the constant $\USp(2n_f)$ rotation
$P$, will leave the meson VEVs invariant.  A straightforward
generalization of these arguments demonstrates that the meson VEVs
with either choice of phase for the nonzero elements do indeed leave a
$U(n_f)$ global symmetry unbroken.

$\hbox{$\bf{m_i=m_0 \neq 0}$}$:
When the hypermultiplet masses are all equal and
nonvanishing, each eigenvalue $\lambda_i$ is one of the two solutions 
(as a
function of $X$) of Eq.~\eqlam, $\lambda_{\pm}=(\mu m_0\pm\sqrt{\mu^2
m_0^2+16\mu X})/4$. Solutions of the full nonlinear, coupled 
equations \eqlam\
and \eqlamX\  may be classified according to the number of 
$\lambda_i$'s which
take on the value $\lambda_+$. Although the total number of solutions 
is the
same, the symmetry breaking patterns are completely different from the
$m_i=0$ case. In vacua where, say $r$ of the $\lambda_i$'s are equal
to $\lambda_+$ (and the remaining are equal to $\lambda_-$), a 
generalization
of the argument above reveals the unbroken
flavor symmetry group to be $U(r)\times U(n_f-r)$ which agrees with 
the classical
analysis of the symmetry breaking patterns in the previous section. 
In fact the
number of such ``$r$-vacua'' also coincides with the semiclassical
prediction. For large enough quark masses $\Lambda_{N=2}\ll m\ll \mu$,
and assuming that $X\ll\mu m^2$ we find from Eq.~\eqlamX\
\eqn\eqmass{X\sim e^{2\pi ik/(n_c-2r-2)} \mu m^{2(n_f-2r)\over 
n_c-2r-2}
\Lambda_{N=2}^{2(n_c-n_f-2)\over{n_c-2r-2}};\quad\quad 
k=1,2,\ldots,n_c-2r-2.}
Note that the multiplicity of solutions, $n_c-2r-2$ is precisely the 
Witten
index for the $SO(n_c-2r)$ SUSY Yang-Mills with $n_c-2r\ge 5$. Note 
also that
the solutions are consistent with the assumption $X\ll \mu m^2$. 
Finally, as
there are $\pmatrix{n_f\cr r}$ ways of picking a vacuum configuration 
with
$r$ nonzero $\lambda_i$'s, the number of vacua with $U(r)\times 
U(n_f-r)$
symmetry is $(n_c-2r-2)\pmatrix{n_f\cr r}$ which again agrees with
semiclassical predictions.

\subsec{Case: $2n_f=n_c-4$}

As described in \isone, the squark VEVs in the $N=1$ theory (in the
$\mu\rightarrow \infty$ limit) break $SO(n_c)$ to $SO(4)\simeq
SU(2)_L\times SU(2)_R$.  The dynamically generated superpotential now
arises due to the combined effect of gaugino condensation in each
independent $SU(2)$ sector.  Since there are two possible choices of
phase for the gluino condensate from each $SU(2)$, one finds two
inequivalent branches in the $N=1$ theory, with superpotential \istwo,
\eqn\supone{W^{N=1}= {1 \over 2} (\epsilon_L + \epsilon_R) \left({16
\Lambda_{n_c, n_c-4}^{2(n_c-1)} \over \det M } \right)^{1/2}}
where $\epsilon_L = \pm1;$ $\epsilon_R = \pm1$.  The branches with $\epsilon_L = \epsilon_R$ give
\eqn\numso{ {\cal N}_1= (n_c-n_f-2) \, 2^{n_f}}
solutions exactly as in the earlier example, with unbroken $U(n_f)$
symmetry in the limit of massless quarks ($m_i=0$). 
However, now we have additional vacua from the two possible
phases with $\epsilon_L=-\epsilon _R$.  The instanton-induced
superpotential vanishes and the tree level terms alone account for the
exact superpotential:
\eqn\veqfive{ W= - {1 \over 2\mu} \Tr (\bJ M \bJ M ) + { 1\over 2} \Tr
{m M }.}
The vacuum expectation value of $M$ is thus {\it uniquely} determined
to be $M={\mu\over 2}\pmatrix{0 & \tilde{m}\cr \tilde{m} & 0}$.  When
the quark masses vanish, the meson condensate simultaneously vanishes,
thus preserving the full $\USp(2n_f)$ symmetry.  Since there are two
choices of phases with $\epsilon_L=-\epsilon_R$ we obtain
\eqn\ntwo {{\cal N}_2=2}
vacua in which the $\USp(2n_f)$ global symmetry remains unbroken in
the $m_i \to 0$ limit.  The total number of $N=1$ vacua ${\cal
N}={\cal N}_1+{\cal N}_2$ is now in perfect agreement with the general
semiclassical formula Eq.~\semic\ which gives in this case
\eqn\semictwo{{\cal N} = (n_c-n_f-2) \; 2^{n_f} + 2,}
the last term being the correction due to the irregularity of the
Witten index for the term with $r=n_f$: $w(n_c-2n_f )=w(4)= 4 $
instead of $n_c-2n_f -2=2$.  Thus the ${\cal N}_1 +2$ vacua found
above from the large $\mu$ analysis precisely match the total number $
{\cal N}$ found by semiclassical methods.

As before, with equal and non-zero hypermultiplet masses the ${\cal
N}_1$ vacua split into groups of ``$r$-vacua'', each with $U(r)\times
U(n_f-r)$ flavor symmetry.  The ${\cal N}_2$ vacua on the other hand
with $M={\mu m_0\over2}\pmatrix{0 & \I1\cr \I1 & 0}$ preserve a
$U(n_f)$ global symmetry in the equal mass case.  They are smoothly
connected to some of the semiclassical vacua with $r=n_f$ and $U(n_f)$
symmetry.

\subsec{Case: $2n_f=n_c-3$}

Now the squark VEVs in the $N=1$ limit (i.e. in the absence of the
adjoint scalar) break the gauge group to $SO(3)\simeq SU(2)$.  It is
useful however, to consider the limit where $2n_f-1$ squark VEVs (or
meson eigenvalues) are taken to be much larger than the remaining one. 
In this regime a description in terms of an $SU(2)_L\times SU(2)_R$
gauge theory with one flavor becomes appropriate.  This flavor obtains
an expectation value which breaks the gauge group to a diagonally
embedded $SU(2)$.  The dynamical superpotential is generated via
instantons both in the diagonal $SU(2)$ and also in the broken parts
of $SU(2)_L$ and $SU(2)_R$.  Once again, the corresponding $N=1$
theory has two inequivalent branches, with superpotential \istwo,
\eqn\casefive{W^{N=1}= 4(1 + \epsilon) {
\Lambda_{n_c,n_c-3}^{2(n_c-1)} \over \det M },}
and $\epsilon=\pm 1.  $ The branch with $\epsilon=1$ yields
\eqn\ds{ {\cal N}_1= (n_c-n_f-2) \, 2^{n_f}}
vacua with $U(n_f)$ symmetry in the limit of massless quarks, and
groups of $r$-vacua when the quarks have equal masses.

The description of the $\epsilon=-1$ branch is incomplete as it
stands.  Upon decoupling one of the flavors it fails to reproduce the
two vacua of the $\epsilon_L = - \epsilon_R $ branch of the
$2n_f=n_c-4$ theory.  Intriligator and Seiberg argued \isone\ that
this apparent conflict can be rectified if there are additional color
singlet, massless particles $q_i$ ($i=1,2,\ldots, 2n_f$), coupled to
the mesons through the superpotential,
\eqn\wtwo{W_2^{N=1}\approx M^{ij} q_i q_j.}
Thus when a flavor $f$ is decoupled by adding a large mass, $m_f
M^{ff}$, the equation of motion for $M^{ff}$ gives two vacua $q^f=\pm
\sqrt {-m_f/2}$, with vanishing superpotential.  Furthermore, the
theory (at $M=0$) with the massless particles $M$ and $q$ satisfies
the 't Hooft anomaly matching conditions \foot{The field $q_i$ was
identified by Intriligator and Seiberg as an exotic composite
$(Q)^{n_c-4}W_\alpha W^\alpha$ which is a glueball for $n_c=4$.}.

The low energy theory at large $\mu$ is described by the superpotential,
\eqn\veqsix{ W= - {1 \over 2\mu} \Tr (\bJ M \bJ M ) + { 1\over 2} \Tr
{m M } + \Tr (q q^T M),}
where $q$ is a $2n_f$ component vector.  The equations of motion for
$q$ and $M$ are
\eqn\eqforq{ M^{ij} q_j=0;\;\;\;- {1 \over \mu} (\bJ M \bJ)+{1\over
2}m + q q^T=0.}
These two equations have a unique solution corresponding to $q_i=0$
and $M=\mu m/2$.  This result may be inferred from the two expressions
obtained by multiplying the second equation by $M$ -- first on the
left and then on the right.  Direct comparison of the two resulting
expressions yields $\bJ Mm\bJ =-mM$.  This in turn implies that
$M=\pmatrix{0 & B\cr B & 0}$ and $[B,\tilde m]=0$, and consequently
$q_i=0$.  So the branch with $\epsilon=-1$ has $M=\mu m/2$, which is
just ${\cal N}_2=1$ additional vacuum that is $\USp(2n_f)$ symmetric
in the $m_i \to 0$ limit (since the meson VEV is proportional to the
mass matrix).  The total number of vacua ${\cal N}=n_f 2^{n_f}+1$
agrees with \semic.  When $m_i=m_0\neq 0$ this additional vacuum with
$M=\mu m/2$ is clearly $U(n_f)$ symmetric and can be smoothly related
to one of the semiclassical $r$-vacua with $r=n_f$ and
$SO(n_c-2n_f)=SO(3)$ unbroken gauge symmetry.

\subsec{Case: $2n_f=n_c-2$}

All the theories with $2n_f\ge n_c-2$ pose an interesting question
that did not arise in the previous examples.  This concerns the
appearance of semiclassical ($m_i$ large) vacua which seem to be in
the Higgs and Coulomb phases.  One may then ask what happens to such
vacua as we tune the quark masses to smaller and smaller values and
semiclassical descriptions cease to hold; i.e. can we identify the
quantum ($m_i$ small/vanishing) description of these vacua in the
large $\mu$ theory?  This question assumes a certain importance in an
$SO(n_c)$ gauge theory with matter in the vector representation
wherein one expects a clear distinction between the confining phase
and a Higgs/Coulomb phase.  The phases can be unambiguously
distinguished by the expectation value of a spinorial Wilson loop,
i.e. the large-distance potential between two electric charges in the
spinor representation of the gauge group.  Since supersymmetry
(holomorphy) disallows a phase transition as we dial $m_i$ from large
to small values, we should expect to find non-confining vacua in
precisely the same phases (Higgs/Coulomb as determined
semiclassically) in the quantum (or small $m_i$) regime as well.  It
should therefore be possible to follow these vacua to the regime of
small $m_i$ and identify the mechanisms in the quantum theory
responsible for confinement, Higgs or Coulomb-like phases.  We will
address this issue in what follows.  Note that there is no such
distinction in $SU(n_c)$ theories with fundamental matter.  Indeed, as
is well-known, the $SU(n_c)$ Seiberg-Witten theories (see \swtwo)
provide examples where one may continuously pass between solitons
(monopoles) of the theory and elementary particles (quarks) and
therefore smoothly interpolate from a confining to a non-confining
(Higgsed) theory.

In the theory with $2n_f=n_c-2$, while there seems to be no
semiclassical Higgs vacuum, there is a single vacuum with
$r=n_f=n_c/2-1$ and $SO(2)\simeq U(1)$ gauge symmetry.  Unlike all the
previous examples and all other ``$r$-vacua'', this vacuum is {\it
non-confining} and in particular is in a Coulomb phase.  Therefore in
the absence of phase transitions we should find a single Coulomb
vacuum in the small mass theory as well.

$\hbox{$\bf{m_i=0}$}$:
It is known from \istwo\ that in the limit where the adjoint $\Phi$ is
decoupled, the low energy $N=1$ theory is broken to $SO(2)\simeq U(1)$
by squark VEVs.  In this case, there is no superpotential for the
mesons -- the theory has a quantum moduli space labelled by the
expectation value of $M_{ij}$.  The effective gauge-coupling $\tau$ of
the photon multiplet in the Coulomb phase can depend only on the
flavor singlet $\det M$, and as usual it is interpreted as the complex
structure of an elliptic curve.  The gauge-coupling $\tau$ is singular
(as a function of $\det M$) at $\det M=0$ and $\det
M=U_1=16\Lambda^{(2n_c-4)}_{n_c,n_c-2}$ where additional light degrees
of freedom appear.  The light particle spectrum can be inferred from
the monodromies resulting from taking $M$ around the singular (or
degeneracy) points of the elliptic curve describing the low energy
Coulomb phase.  One finds that near $\det M=U_1$ a pair of light
monopoles $E^\pm$ with magnetic charges $\pm 1$ appear while at $\det
M=0$, $2n_f$ pairs of monopoles $(q_i^+,q_i^-)$ with magnetic charges
$\pm1$ become light.  In the vicinity of $\det M =U_1$ the monopoles
$E^\pm$ are described by a superpotential, $W^{N=1}=(\det M - U_1)
E^{+} E^{-}$.  In the large $\mu$ theory this leads to an
effective superpotential of the form
\eqn\supuone{ W= (\det M - U_1) E^{+} E^{-} - {1 \over 2\mu} \Tr (\bJ
M \bJ M ) + { 1\over 2} \Tr {m M }; \quad U_1=16 \Lambda_{n_c,
n_c-2}^{2 n_f}.}
It must be emphasized that this superpotential is expected to be valid
only near $\det M \approx U_1$.  The equations of motion following
from it are
\eqn\equone{(\det M) E^{+} E^{-} {\I1} - {1 \over \mu} \bJ M \bJ M +
{1\over 2}m M =0,}
and
\eqn\eqsuone{ (\det M - U_1) E^{+} = (\det M - U_1) E^{-}=0.}
As in the theories with $2n_f\le n_c-5$, it can be readily shown that
$M=\pmatrix{0 & B \cr B & 0}$ and $B={\rm diag}
(\lambda_1,\lambda_2,\ldots,\lambda_{n_f})$, where the $\lambda_i$'s
satisfy
\eqn\lame{ -{1 \over \mu } \lambda_i^2 + {1\over 2}m_i \lambda_i +
X=0; \quad \quad \det M=\prod_{i=1}^{n_f} \lambda_i^2 = U_1;}
with $X=E^+E^-U_1$.
The magnitudes of $E^+$ and $E^-$ are related as usual by the D-term
condition, $|E^+|=|E^-|$.  The phase of $E^-$ (say) can always be
removed by a (dual) $U(1)$ gauge rotation, while that of $E^+$ is
then fixed by the condition $E^+E^-=X/U_1$.  In the $m_i\rightarrow 0$
limit, then
\eqn\ans{\lambda_i=\pm\sqrt{\mu X};\quad{\rm and}\quad
X=U_1^{1/n_f}e^{2\pi ik/n_f};\quad k=1,2,\ldots,n_f.}
These solutions therefore yield $n_f 2^{n_f}$ possible vacuum
configurations.  These meson VEVs also preserve a $U(n_f)$ flavor
symmetry -- arguments for this proceed exactly as in subsection (3.1);
we will not repeat them.  Note that all these solutions satisfy $\det
M=U_1$ and hence the use of \supuone\ is justified.  There are also
``fake'' solutions, $E=0,\,\, M=0$, and $E=0,\,\, M=\mu m/2$, which
however lie far from the point $\det M = 16 \Lambda_{n_c, n_c-2}^{4
n_f}$ and must therefore be discarded.  Hence we have found a total of
${\cal N}_1$ vacua with $U(n_f)$ symmetry in the $m_i\rightarrow 0$
limit:
\eqn\casel{{\cal N}_1=n_f 2^{n_f}=(n_c-n_f-2)2^{n_f}.}

Turning to the singularity at $\det M=0$, the light degrees of freedom
near this point, namely the mesons and the $2n_f$ light monopoles are
described by the effective superpotential,
\eqn\veqseven{ W= - {1 \over 2\mu} \Tr (\bJ M \bJ M ) + { 1\over 2}
\Tr {m M } + f(\det M)\, q_i^+ q_j^- M^{ij}.}
$f$ is an undetermined holomorphic function of $\det M$ satisfying
$f(0)=1$ so that near $\det M\approx 0$ we may write $f\approx 1+
t\det M$.
The equations of motion are
\eqn\veqeight{ - {1 \over \mu} (\bJ M \bJ ) + {1\over 2}m + {1\over 2}
[q^+ q^{-T} + q^- q^{+T}] =0,}
and
\eqn\orthog{ q_j^- M^{ij}= 0, \qquad q_i^+ M^{ij}=0.}
Following the arguments in the case of the theory with $2n_f=n_c-3$,
the meson condensate is found to be $M=\pmatrix{0 & B\cr B & 0}$ where
$B$ is diagonal; ${B=\rm diag}(\lambda_1,\lambda_2,\ldots,\lambda_{n_f})$.  However, unlike
the previous case, in addition to the vacuum with $q^\pm=0$ and $M=\mu
m/2$, there are also vacua where monopoles condense.  To see this, let
us write $q^\pm=\pmatrix{a^\pm \cr b^\pm}$ where $a^\pm_i, b^\pm_i$
are $n_f$-component vectors.  Now suppose that $a_1^+\neq 0$.  This
along with the equations of motion \veqeight\ and \orthog, then
automatically implies that except for $a_1^+$ and $b_1^-$ all other
monopole fields must vanish and that $\lambda_1=0$.  Furthermore, the
D-term condition leads to $|a_1^+|=|b_1^-|$.  The phase of $a_1^+$ can
be fixed by a $U(1)$ gauge rotation and then the phase of $b_1^-$ is
uniquely determined by the equations of motion, namely
\eqn\veqnine{a_1^+b_1^-+{1\over 2}m_1=0;\quad\quad|a_1^+|=|b_1^-|;
\quad\quad B={\mu\over 2}{\rm diag}(0,m_2,\ldots,m_{n_f}).}
In the massless limit, these meson VEVs preserve $\USp(2n_f)$
symmetry.  Since the above arguments would apply equally well if we
chose any one of the $q_i^+$'s to be non-zero, we have found $2n_f$
new solutions where monopoles condense and the meson VEVs preserve the
full $\USp(2n_f)$ global symmetry (for massless quarks).  Thus,
including the vacuum with $q^\pm=0$ and $M=\mu m/2$ we get ${\cal
N}_2=2n_f+1$ vacua with unbroken flavor symmetry when the quark masses
are zero.  The total number of vacua is ${\cal N}_1+{\cal N}_2
=(n_c-n_f-2)2^{n_f}+2n_f+1$ which is indeed the correct number.

$\hbox{$\bf{m_i=m_0\neq 0}$}$:
when $m_i=m_0\neq 0$, the $2n_f$ vacua with monopole $q_i^\pm$ condensation have
a $U(1)\times U(n_f-1)$ symmetry and correspond to a half of the semiclassical vacuum states with $r=n_f-1$ and $SO(4)$ gauge symmetry. 
The first group of ${\cal N}_1$ ground states can be shown as before to split into groups of $r$-vacua with $U(r)\times U(n_f-r)$ global symmetry, all in confined phase. The subset with $r=1$ corresponds to the other half of the semiclassical vacuum states with $SO(4)$ gauge symmetry.

The remaining vacuum, with $M=\mu m/2$ and $q^\pm=0$ preserves a $U(n_f)$ global symmetry group.  The absence of monopole VEVs indicates that the theory in this vacuum is non-confining and in fact continues to be in the Coulomb phase rather than an IR free phase, since all monopoles are massive.  
Hence, we have found a unique $U(n_f)$-symmetric vacuum in the Coulomb phase which, as we dial the quark masses $m_0$, must smoothly transform into the unique $r=n_f$ semiclassical vacuum with $SO(2)\simeq U(1)$ gauge symmetry.
In particular, the quantum (or small $m$) description consists of dual photons that couple locally to the (massive) monopoles -- however the potential between two electric test charges at a separation $R$ still behaves as $1/R$ corresponding to the usual Coulomb phase.

\subsec{Case: $2n_f=n_c-1$}

In the infrared, the corresponding $N=1$ theory is dual to a magnetic
$SO(3)$ gauge theory with $2n_f$ flavors of dual quarks and singlet
mesons $M_{ij}$.  Since $2n_f\ge 5$, this theory is infrared free.  In
the presence of the heavy (but finite mass) adjoint $\Phi$ and in the
presence of quark masses, the superpotential (near $M=0$) turns out to
be
\eqn\supnine{ W= { 1\over 2\kappa} M_{ij} q_i \cdot q_j - { \det M
\over 64 \Lambda^{2 n_c-5}_{n_c, n_c-1}} - {1 \over 2\mu} \Tr (\bJ M
\bJ M ) + { 1\over 2} \Tr {m M }.}
$\kappa$ is a dimensionful normalization scale (we assume that
$\kappa\sim\Lambda_{n_c,n_c-1}$) which relates the mesons $M$ of the
electric theory to the corresponding dimension one operators in the
magnetic description via $M_m=M/\kappa$.  The dynamical scales
$\Lambda_{n_c,n_c-1}$ and $\tilde{\Lambda}_{3,n_c-1}$ of the electric
and magnetic descriptions respectively and $\kappa$ are related via
\eqn\kappeq{\Lambda^{2n_c-5}_{n_c,n_c-1}
\tilde{\Lambda}_{3,n_c-1}^{4-n_c}\sim \kappa^{n_c-1}.}

As before, we will analyze the vacuum structure of the theory when the
quark masses $m_i$ are non-zero and then focus on the special cases
where the masses are all equal, and also when they are vanishingly
small compared to the $N=2$ dynamical scale.  For generic, non-zero
quark masses the mesons will condense and render the dual quarks
massive.  We will classify the vacua according to the rank of the
meson VEVs.

When rank$(M)=2n_f$, all the dual quarks are massive and can be
integrated out, leaving behind a pure, magnetic $SO(3)$ gauge theory
wherein a superpotential is generated as usual via gaugino
condensation.  We remark that the equations of motion for the dual
quarks, following from \supnine, namely $M_{ij}q^a_j=0$ ensure that
when rank$(M)=2n_f$ the magnetic squarks have vanishing VEVs and so
the dual $SO(3)$ symmetry is unbroken.  The low energy, pure gauge,
magnetic $SO(3)$ theory then has a scale
$\tilde{\Lambda}^3_{3,0}=\tilde{\Lambda}^{4-n_c}_{3,n_c-1}\det(M/\kappa)$
and has two branches, each labelled by the phase of the gluino
condensate.  The contribution to the superpotential from instanton
effects in the magnetic theory is thus $2\epsilon
\tilde{\Lambda}^{4-n_c}_{3,n_c-1}\det(M/\kappa)$ with $\epsilon=\pm1$
labelling the two possible phases of the gaugino condensate.  Using
Eqs.~\supnine\ and \kappeq\ we find that the $\epsilon=+1$ branch has
a superpotential given by
\eqn\pls{W_+=-{1 \over 2\mu} \Tr (\bJ M \bJ M ) + { 1\over 2} \Tr {m M
},}
while the $\epsilon=-1$ branch develops a superpotential $W_-$:
\eqn\mins{W_-=- { \det M \over 32 \Lambda^{2 n_c-5}_{n_c, n_c-1}} - {1
\over 2\mu} \Tr (\bJ M \bJ M ) + { 1\over 2} \Tr {m M }.}

The $\epsilon =+1$ branch yields a single vacuum with
\eqn\epsvev{\langle M\rangle=\mu m/2={\mu\over 2}\pmatrix{0 & \tilde
m \cr \tilde m & 0}.}

At this point we should clarify the various mass scales involved. 
Since the meson VEVs $\sim \mu m$, the dual quarks have masses
$\sim\mu m/\kappa$.  On the other hand the meson masses
$\sim\kappa^2/\mu$.  Since we have chosen to retain the mesons as the
light degrees of freedom, we must have $\kappa^2/\mu\ll {\mu
m/\kappa}$.  which in turn means that
\eqn\scales{ \left( {\Lambda_{N=2}\over\mu} \right)^{n-4\over 2n-5}\ll
{m\over\Lambda_{N=2}},}
if we assume that $\kappa\sim\Lambda_{n_c,n_c-1}$.  Within this range
of parameters (even if we assume $m/\Lambda_{N=2}\ll 1$), we expect
our analysis to be valid.

Interesting conclusions may be drawn from the form of the meson VEVs
\epsvev\ in this vacuum.  Firstly, since $\epsilon$ is the phase of
the gluino condensate, it may be thought of as the theta angle of the
dual theory.  Thus the $\epsilon=+1$ branch corresponds to the
confinement of dual quarks while $\epsilon=-1$ (changing theta by
$\pi$) corresponds to dyonic or oblique confinement in the dual
theory.  Confinement of the magnetic degrees of freedom in the
$\epsilon=+1$ vacuum can be interpreted as condensation of electric
charges of the original theory.  Secondly, the meson VEV \epsvev\
coincides precisely with the classical prediction for the meson
condensates in the $r$-vacuum with $r={n_c-1\over 2}$ which,
semiclassically corresponds to a Higgs phase.  Finally, since the
instanton effects disappear when $\epsilon=+1$, the classical
superpotential does not receive any quantum corrections.  It is
therefore reasonable to conclude that the single $\epsilon=+1$ vacuum in
the dual theory can be smoothly connected to the single semiclassical
vacuum in the Higgs phase in the original description, while
preserving the distinction between confining and non-confining phases
in $SO(n_c)$ gauge theories without spinorial matter fields.  When the
quark masses are taken to be vanishingly small, this vacuum has an
unbroken $\USp(2n_f)$ flavor symmetry, while the equal mass limit
$m_i=m_0\neq 0$ leaves a $U(n_f)$ symmetry.  These patterns of flavor
symmetry breaking also agree with the classical analysis of the
$r=n_f=(n_c-1)/2$ theory where the gauge group appears to be
completely Higgsed.

Using the superpotential $W_-$ \mins, the meson VEVs in the $\epsilon
=-1$ branch can be shown to have the form $M=\pmatrix{0 & B\cr B & 0}$
with $B={\rm diag}(\lambda_1,\lambda_2,\ldots,\lambda_{n_f})$.

For $m_i\ll \Lambda_{N=2}$ we find that there is one solution with
\eqn\smallvev{\lambda_i\approx\mu{m_i\over 2} \left[ 1+O \left(
\left({ m \over \Lambda_{N=2}} \right)^{n_c-3} \right) \right],}
which preserves the full $\USp(2n_f)$ symmetry in the massless limit,
while leaving a $U(n_f)$-symmetric vacuum in the equal (non-vanishing)
mass case.

The $\epsilon=-1$ branch also yields a whole other set of vacua with
{\it non-vanishing} meson VEVs in the $m_i\to 0$ limit with
\eqn\largevevs{\lambda_i=\pm\sqrt {\mu X}; \quad X= -{\det M\over
32\Lambda^{2n_c-5}_{n_c,n_c-1}}\sim e^{2\pi
ik/(n_f-1)}\mu\Lambda_{N=2}^2,\quad k=1,2,\ldots, n_f-1,}
and these yield a total of
\eqn\nonetwo{{\cal N}_1=(n_f-1)2^{n_f}}
vacua.  The global group is broken to $U(n_f)$ in the massless limit
in all these vacua, while in the equal mass case they reproduce the
groups of $r$-vacua with $U(r)\times U(n_f-r)$ flavor symmetry.

We now look for possible ground states with rank$(M)<2n_f$.  We find
that supersymmetric ground states exist only when rank$(M)=2n_f-2$. 
Below the squark-mass scale set by the meson expectation values $\sim
\langle M\rangle/\kappa$ the theory looks like a magnetic $SO(3)$
gauge theory with two light dual quarks.  The $SO(3)$ gauge theory
with two light flavors is known to be either in a Higgs phase or in an
IR free phase and hence there can be no further dynamical corrections
to the superpotential.  In fact the equations of motion from \supnine\
indicate dual squark-condensation and consequent Higgsing of the dual
gauge group.  We find that $M=\pmatrix{0 & B \cr B & 0}$ , where
\eqn\rank{B={\mu\over 2} \, {\rm diag}(0,m_2,m_3,\ldots),}
and, denoting $q_{i+n_f}$ by $\tilde{q}_i$:
\eqn\example{q_1= \pmatrix{d \cr i d \cr 0 }, \quad {\tilde q}_1=
\pmatrix{d \cr -i d \cr 0 },}
so that
\eqn\exampletwo{ q_1 \cdot q_1={\tilde q}_1 \cdot {\tilde q}_1=0;
\quad {\tilde q}_1 \cdot q_1= 2 d^2= -\kappa m_1.}
Note that the relative magnitudes of $ {\tilde q}_1 $ and $ q_1$ are
fixed by the D-term constraint.  Also, the overall signs can be
brought in the above form (with $d=+\sqrt {-m_1\kappa/2}$) by
combining dual $SO(3)$ gauge and global $SO(2 n_f) \subset \USp(2n_f)$
rotations.  Analogous solutions can be constructed by choosing any one
of the $n_f$ pairs of squarks to be nonvanishing so that there are
$n_f$ solutions of this type, all $\USp(2n_f)$ symmetric.  Therefore,
including the single Higgs vacuum we have found a total of
\eqn\ntwotot{{\cal N}_2=n_f+2} vacua with vanishing VEVs and
$\USp(2n_f)$ symmetry in the theory with massless quarks.

The total number of vacua is then ${\cal N}={\cal N}_1+{\cal
N}_2=(n_f-1)2^{n_f}+n_f+2$ which matches the semiclassical result in
Eq.~\semic.  We remark that we have not found any Coulomb vacua in the
large-$\mu$ theory, which is also completely consistent with classical
expectations.

\subsec{Case: $2n_f=n_c$}

These theories have a dual description in terms of an $SO(4)\simeq
SU(2)_L\times SU(2)_R$ gauge theory with $2n_f$ flavors of dual quarks
in the $(\bf 2,\bf 2)$ representation, and $SO(4)$ singlet mesons with
superpotential
\eqn\supten{W^{N=1}={ 1\over 2\kappa} M_{ij} q_i \cdot q_j,}
where $\kappa$ is related to the dynamical scales of the electric and
magnetic descriptions through
$\Lambda^{2n_c-6}_{n_c,n_c}\tilde{\Lambda}_{n_c} \sim\kappa^{n_c}.$
The complete low-energy, effective superpotential for our theory is
simply
\eqn\supfull{ W= { 1\over 2\kappa} M_{ij} q_i \cdot q_j - {1 \over
2\mu} \Tr (\bJ M \bJ M ) + { 1\over 2} \Tr {m M }.}
Proceeding as in the previous case, we classify vacuum states
according to the rank of the meson VEVs.

When rank$(M)=2n_f$ we integrate out the dual quarks to obtain a
gaugino-condensate-induced superpotential in the pure, magnetic
$SO(4)$ theory.  Since the dual $SO(4)$ decomposes into
two-independent $SU(2)$'s, we must have two inequivalent branches, one
with no dynamically generated superpotential.  The dynamical
contribution to the superpotential in these branches is
$=2(\epsilon_L+\epsilon_R)\tilde \Lambda^3$ where $\tilde\Lambda$
denotes the scale of either one of the $SU(2)$'s contained in the dual
$SO(4)$, and $\epsilon_L,\epsilon_R$ are the phases of
$\langle\lambda\lambda\rangle_{L,R}$ in the respective $SU(2)$
sectors.

The branch with $\epsilon_L\epsilon_R=-1$ results in an effective
theory with no instanton contributions to the superpotential which
therefore is purely classical:
\eqn\branchcl{W(\epsilon_L\epsilon_R=-1)=- {1 \over 2\mu} \Tr (\bJ M
\bJ M ) + { 1\over 2} \Tr {m M }.}
and yields two vacua (since there are two possible choices of phase
with $\epsilon_L+\epsilon_R=0$) with $M=\mu m/2$.  In the equal mass
theory this preserves a $U(n_f)$ symmetry while in the limit of
vanishing masses the full $\USp(2n_f)$ flavor group is restored.  The
form of the meson VEV, the number of vacua, the flavor symmetry
patterns and finally, the absence of quantum corrections to the
superpotential in these vacua with $\epsilon_L\epsilon_R=-1$, strongly
suggest that these are in fact the large-$\mu$, small $m$ counterparts
of the two Higgs vacua which appear in the semiclassical description
of the original electric theory.  This is actually in perfect accord
with our knowledge of the vacuum structure of $SU(2)_1\times SU(2)_2$
gauge theory with two massless flavors in the $(\bf 2,\bf 2)$
representation.  The theory is known to have a moduli space of vacua
with two singular submanifolds at which magnetic and dyonic degrees of
freedom become light.  Upon turning on masses for the flavors, each
singular submanifold gives two vacua where monopoles (or dyons)
condense.  In the dual theory the two vacua associated with dual quark
confinement (and not oblique confinement) must correspond to the
Higgsed ground states of the electric theory.

The branch with $\epsilon_L=\epsilon_R$ leads to a superpotential with
a form similar to what we have encountered in several previous
examples:
\eqn\branchq{W(\epsilon_L\epsilon_R=1)=
-(\epsilon_L+\epsilon_R){\sqrt{\det M}\over
8\Lambda^{n_c-3}_{n_c,n_c}} -{1 \over 2\mu} \Tr (\bJ M \bJ M ) + {
1\over 2} \Tr {m M }.}
This leads to two vacua with VEVs proportional to the quark masses in
the small mass regime and ${{\cal N}_1=(n_f-2)2^{n_f}}$ vacua with
finite VEVs.  The two vacua with vanishing VEVs (and consequently
unbroken flavor group) have the usual off-diagonal form for the meson
VEVs with $\lambda_i\approx\mu m/2(1\pm O(m/\Lambda_{N=2})^{n_f-2})$. 
The finite VEV vacua which dynamically break flavor to $U(n_f)$ can be
found as in many previous examples.

Now we turn to the cases with rank$(M)<2n_f$ and we find two sets of
supersymmetric ground states, for rank$(M)=2n_f-4$ and for
rank$(M)=2n_f-2$.  When rank$(M)=2n_f-4$, at long distances the theory
looks like a magnetic $SO(4)$ theory with four flavors of dual quarks. 
The corresponding $N=1$ theory (studied by Intriligator and Seiberg)
is at a non-trivial fixed point of the beta function.  In our theory,
in fact using the superpotential \supfull\ we find squark and meson
VEVs of the form
\eqn\lorank {M=\pmatrix{0 & B\cr B & 0};\quad B={\mu\over 2}{\rm diag}
(0,0,m_3,m_4,\ldots,m_{n_f});} \eqn\loranktwo{q_1= \pmatrix{d_1 \cr i
d_1 \cr 0 \cr 0 }, \quad {\tilde q}_1= \pmatrix{d_1 \cr -i d_1 \cr 0
\cr 0 }, \quad q_2= \pmatrix{0 \cr 0 \cr d_2 \cr i d_2 }, \quad
{\tilde q}_2= \pmatrix{ 0 \cr 0 \cr d_2 \cr -i d_2 }, \quad }
where
\eqn\factortwo{ d_i= +\sqrt{-m_1\kappa/2}; \quad d_2= \pm
\sqrt{-m_2\kappa/2}.}
These VEVs break the dual gauge symmetry completely, and should
correspond to confining vacua in the electric description.  Since we
could have chosen the non-vanishing flavors in $\pmatrix{n_f\cr 2}$
ways, there are $2{}_{n_f}\!C_2$ vacua \foot{The factor $2$ in front
of $ {}_{n_f}\!C_2$ is due to the sign choices in Eq.~\factortwo\
which cannot be undone by gauge or global transformations, and hence
are inequivalent.} where the dual gauge group is completely Higgsed
and the flavor symmetry group in the massless limit remains unbroken,
while in the limit of equal masses the flavor group appears to be
$U(2)\times U(n_f-2)$.

When rank$(M)=2n_f-2$ the low energy description is that of the
magnetic $SO(4)$ theory with two quark flavors which is known to be in
the Coulomb phase.  This is confirmed by the solution to the equations
of motion from \supfull\ which lead to dual squark condensates that
break $SO(4)$ to $SO(2)\simeq U(1)$ and meson VEVs are given by $B={\mu\over
2}{\rm diag} (0,m_2,m_3,\ldots)$.  There are $n_f$ such solutions,
which are clearly the large-$\mu$ (dual) descriptions of the $n_f$
Coulomb vacua that are apparent in the semiclassical limit.

Therefore the number of vacua with vanishing VEVs is 
\eqn\ntwop{ {\cal N}_2 =4 + \, n_f + 2 \, {}_{n_f}\!C_2,}
while the total number of
supersymmetric vacuum states is 
\eqn\totno{{\cal N} ={\cal N}_1+{\cal N}_2 =
(n_f-2)2^{n_f}+4+n_f+{}_{n_f}\!C_2}
which agrees with Eq.~\semic.

\subsec{Case: $2n_f>n_c$}

The corresponding $N=1$ theories (when $\Phi$ is decoupled) are dual
to the magnetic $SO(\tilde{n}_c)\equiv SO(2n_f-n_c+4)$ gauge theory
with $2n_f$ dual squarks and singlet mesons $M_{ij}$ with the
effective superpotential
\eqn\dualsup{W={1\over 2\kappa}M_{ij}q_i\cdot q_j
-{1\over 2\mu}\Tr(\bJ M\bJ M)+{1\over 2}\Tr(mM).}
As before $\kappa$ relates the scales of the electric and magnetic 
theories via
\eqn\kappy{\Lambda_{n_c,2n_f}^{3(n_c-2)-2n_f}
\tilde{\Lambda}_{2n_f-n_c+4,2n_f}^{3(2n_f-n_c+2)-2n_f}\sim\kappa^{2n_f}.}

The vacuum structure of this theory can be easily inferred by looking
at the solutions to the classical equations that follow from the
superpotential \dualsup.  These solutions are of the form:
\eqn\clasisol{q_1 = \pmatrix{d_1 \cr i d_1 \cr 0 \cr 0 \cr 0 \cr
\vdots }, \quad
{\tilde q}_1= \pmatrix{d_1 \cr -i d_1 \cr 0 \cr 0 \cr 0 \cr \vdots },
\quad
q_2= \pmatrix{0 \cr 0 \cr d_2 \cr i d_2 \cr 0 \cr \vdots }, \quad
{\tilde q}_2= \pmatrix{ 0 \cr 0 \cr d_2 \cr -i d_2 \cr 0 \cr \vdots },
\quad {\rm etc.},}
so  that
\eqn\classi{ {\tilde q}_i \cdot q_i= 2 d_i^2 =m_i\kappa, \quad
i=0,1,2, \ldots r,}
while all other scalar products among $q$'s and ${\tilde q}$'s are
zero, and
\eqn\clasimes{ A=C=0; \quad B= {\mu\over 2}\, {\hbox {\rm diag}}\, (\,
\underbrace{0, 0,\ldots, 0}_{r}, m_{r+1}, \ldots, m_{n_f}).}
Evidently, ${}_{n_f}C_r$ solutions of this type can be constructed by
choosing any $r$ pairs of dual squarks to have non-vanishing
expectation values.  The solution also indicates that a
$SO(\tilde{n}_c-2r)$ gauge symmetry is left unbroken and therefore for
every $r$ one obtains $w(\tilde{n}_c-2r) {}_{n_f}C_r$ vacuum states
with $\USp(2n_f)$ symmetry in the limit of vanishing masses.  Finally,
when $n_c$ is even, an additional set of ${}_{n_f}C_{\tilde{n}_c/2}$
vacua appear \foot{This point was explained in detail in section 2.}
so that there will be a total of
\eqn\dualntwo{{\cal N}_2=\sum_r^{[\tilde{n}_c/2]}
w(\tilde{n}_c-2r){}_{n_f}C_r+{n_f}C_{\tilde{n}_c}/2}
vacua with small VEVs.  The symmetry breaking patterns implied by the
meson condensates \clasimes\ for equal and nonvanishing masses, are
also in complete agreement with the semiclassical predictions.

Strictly speaking however, these solutions provide only a qualitative
picture of what happens in these vacua.  The VEVs and solutions in
Eqs.  \clasimes\ will receive quantum corrections.  A quantitatively
correct description of the physics can be obtained only after the
dynamical contributions to the superpotential are taken into account. 
Without going into too much detail, we discuss these aspects of the
dynamics briefly below.  As indicated by the ``classical'' solutions
described above, the vacua must be classified according to the rank of
the meson VEVs.  When rank$(M)=2n_f-2r$, with $2r<\tilde{n}_c-4$, the
low energy theory is the magnetic $SO(\tilde {n_c})$ theory with $2r$
dual quarks.  This theory obtains a dynamical superpotential similar
to that encountered in Eq.~\dynamic.  The couplings of the dual quarks
to the mesons $M_{ij}$ ensure that they all get small masses and lead
to $\tilde{n}_c-2r-2$ supersymmetric ground states (and no runaway
vacuum solutions).

When $n_c$ is even and $2r=\tilde{n}_c-4$, the magnetic theory flows
to an $SO(4)\simeq SU(2)\times SU(2)$ theory with two branches as
usual, where dual quarks are confined.  In two of these vacua, the
effective theta angle of the dual theory is zero and the confinement
of the magnetic degrees of freedom may be interpreted as condensation
of the original electric degrees of freedom (and not dyon
condensation).  Taking into account that there are
${}_{n_f}C_{\tilde{n}_c/2-2}$ distinct solutions with
rank$(M)=2n_f-\tilde{n}_c+4$, the total number of Higgs-like vacua in
this description is therefore
$2{}_{n_f}C_{\tilde{n}_c/2-2}=2{}_{n_f}C_{n_c/2}$ which precisely
matches the semiclassical counting of Higgs vacua.  For reasons that
were discussed earlier, the dynamical contributions to the
superpotential cancel out in these vacua.  Thus the solutions
\clasimes\ for $r=\tilde{n}_c/2-2$ are in fact the exact ones and the
meson VEVs coincide with semiclassical values of the condensates in
the semiclassical Higgs vacuum.  For equal masses, $m_i=m_0\neq 0$, in
both the limits a $U(n_f-n_c/2)\times U(n_c/2)$ flavor symmetry is
preserved.

For even $n_c$ the dual theory can also be in a Coulomb phase.  This
occurs when $r={\tilde n}_c/2-1$.  The combinatorial factor determining
the multiplicity of these vacua is ${}_{n_f}C_{\tilde{n}_c/2-1}$ which
is also equal to ${}_{n_f}C_{n_c/2-1}$, the number of semiclassical
Coulomb vacua.

When $n_c$ is odd and $2r=\tilde{n}_c-3$, the dual theory flows to an
$SO(3)$ gauge theory which has two branches, one without a quantum
superpotential.  This branch yields
${}_{n_f}C_{\tilde{n}_c/2-3/2}={}_{n_f}C_{[n_c/2]} $ vacua with a
vanishing effective theta angle and magnetic confinement.  Once again
the meson VEVs given by Eq.~\clasimes\ for this case are the exact
results and are in complete agreement with the semiclassical results
for the Higgs vacuum with a $U([{n_c\over 2}])\times U(n_f-[{n_c\over
2}])$ global symmetry.

Having completely classified and clarified the ${\cal N}_2$ vacua with
small VEVs, let us turn to the finite VEV vacua where the flavor group
is expected to be broken dynamically to $U(n_f)$.  Such vacua must
have rank$(M)=2n_f$ rendering all the dual quarks massive.  The dual
quarks must be integrated out giving a low-energy, pure $SO(\tilde
n_c)$ gauge theory with gaugino-condensation.  The dynamical scale of
this pure superglue theory is obtained as usual by a one-loop matching
and results in the following superpotential:
\eqn\dualsup{\eqalign{W=&{1\over 2}(n_c-2n_f-2)\left({\det
M}\over{16\Lambda_{n_c,2n_f}^{3n_c-2n_f-6}}\right)^{1/(2n_f-n_c+2)}\cr
&-{1\over 2\mu}\Tr(\bJ M\bJ M)+{1\over 2}\Tr(mM).}}
Note that this is just a continuation of \dynamic\ to the large flavor
case.  Solutions of the equations of motion are of two types.  The
first type is a $\USp(2n_f)$-symmetric set of $\tilde
n_c-2=2n_f-n_c+2$ vacua which has already been qualitatively discussed
earlier.  These vacua are characterized by the meson eigenvalues
$\lambda_i=\mu m_i (1+e^{2\pi ik/(\tilde
n_c-2)}O((m/\Lambda_{N=2})^{2n_c-2n_f-4}))$.

Ground states with dynamical flavor-breaking can be located in exactly
the same way as before, with a multiplicity,
\eqn\allnone{{\cal N}_1=(n_c-2n_f-2)2^{n_f}.}
The meson VEVs in this group of vacua is also proportional to $\mu
\Lambda_{N=2}$.

The discussion of the equal mass case $m_i=m_0\neq 0$ also proceeds in
the same way as before.  While the first category of ${\cal N}_1$
theories yields groups of $r-$vacua with $U(r)\times U(n_f-r)$ flavor
symmetry, the second group of vacua have semiclassical VEVs (Eqs. 
\clasisol\ and \clasimes)which also preserve $U(r)\times U(n_f-r)$
flavor group in the equal mass case.

Finally, we turn to the enumeration of all the quantum vacua and
comparison with the semiclassical formula \semic.  Let us first assume
that $\tilde n_c$ is even.  Then we have
\eqn\evenntwo{\eqalign{{\cal N}_2=&\sum_r^{\tilde n_c/2}w(\tilde
n_c-2r){}_{n_f}C_r +{}_{n_f}C_{\tilde n_c/2} =\sum_r^{\tilde
n_c/2-3}(\tilde n_c-2r-2){}_{n_f}C_r +\cr &2{}_{n_f}C_{\tilde
n_c/2}+{}_{n_f}C_{\tilde n_c/2-1} +4{}_{n_f}C_{\tilde n_c/2-2}.}}
A change of variable $r\to n_f-r$ allows us to rewrite the above as
\eqn\ntwore{{\sum}_{r=0}^{n_c/2}w(n_c-2r){}_{n_f}C_r+
{}_{n_f}C_{n_c/2}-{\cal N}_1,}
and hence ${\cal N}_1+{\cal N}_2$ agrees with the semiclassical result \semic.  Note that the terms ${}_{n_f}C_{n_c/2}$ and ${}_{n_f}C_{\tilde n_c/2}$ appear only when $n_c$ is even.  When $n_c$ is odd the number of $\USp(2n_f)-$symmetric vacua is
\eqn\oddntwo{\eqalign{{\cal N}_2=&\sum_r^{(\tilde n_c-1)/2}w(\tilde n_c-2r){}_{n_f}C_r =\sum_r^{(\tilde n_c-1)/2-2}(\tilde n_c-2r-2){}_{n_f}C_r +\cr &{}_{n_f}C_{(\tilde n_c-1)/2}+2{}_{n_f}C_{(\tilde n_c-1)/2-1}.}}
Changing the variable $r\to n_f-r$ we find
\eqn\oddntwore{{\sum}_{r=0}^{(n_c-1)/2}w(n_c-2r){}_{n_f}C_r-{\cal N}_1,}
which is exactly what we expect from the classical analysis.

%: The effective descriptions of the small $\mu$ theory
\newsec{The effective descriptions of the small $\mu$ theory}
In this section we attempt to provide a description of the theory in the limit of {\it small $\mu$} and {\it small $m_i$}.  This is the regime in which $N=2$ supersymmetry is only softly broken.  The low-energy degrees of freedom in some of the $N=1$ vacua were actually identified by Argyres, Plesser and Shapere (APS) in \arplsha.  Below, we briefly review the arguments of APS presented in \arplsha.  We also identify an important ingredient of the physics that was missing in the analysis by these authors.  Counting the vacua and keeping track of the flavor symmetries therein provides a powerful check on the resulting picture.

As described in detail in \arplsha\ the moduli space of $N=2$ $SO(n_c)$ SUSY-QCD consists of Coulomb $(\Phi\neq 0, Q=0)$, Higgs $(\Phi=0, Q\neq 0)$ and mixed $(\Phi\neq 0, Q\neq 0)$ branches.  The Higgs and mixed branch `roots' (i.e. where they meet the Coulomb branch) are singular submanifolds with massless hypermultiplets and enhanced gauge symmetry respectively.  The mixed branch roots have $SO(r)\times U(1)^{(n_c-r)/2}$ gauge symmetry with $n_f$ massless flavors and $n_c-r$ even.  Since they exist only for $0\leq r \leq n_f$, the theories at the roots are all IR free.  Upon breaking SUSY to $N=1$, these singular submanifolds get completely lifted with the exception of isolated points where the corresponding Seiberg-Witten curves are maximally degenerate -- that is, where $(n_c-r)/2$ mutually
local monopole hypermultiplets $\{e_k,\tilde{e}_k\}$ charged under the
$U(1)$'s become massless.  In fact there is precisely one such point at
the $r-$branch roots where this occurs, and has $r=2n_f-n_c+4$.  Thus
Seiberg's dual gauge group $SO(2n_f-n_c+4)$ \seiberg\ makes a natural
appearance, although this does not by itself constitute a proof or
derivation of $N=1$ duality.  In fact the actual story, as we
describe below, is perhaps more complex.
\bigskip
%: Table 3
\begintable
  | $SO(2n_f-n_c+4)$  | $U(1)_1$ | $\cdots$ | $U(1)_{n_c-2-n_f}$ \cr
 $n_f\times Q$ | $\bf 2n_f{-}n_c{+}4$ |  $0$ | $\cdots$  | $0$ \cr
 $e_1$  | $\bf 1$ | $1$|$\cdots$  |$0$  \cr
 $\vdots$ |$\vdots$ |$\vdots$ |$\ddots$ | $\vdots$ \cr
 $e_{n_c-2-n_f}$|$\bf 1$ |$0$|$\cdots$|$1$
\endtable~{\bf Table 3:} Charges of light degrees of freedom at the special point.
\bigskip

Choosing a basis where each light monopole multiplet $e_k$ at the special point has charge $1$ under one of the Coulomb factors $U(1)_k$, we may summarize the charges of the light degrees of freedom
as in Table 3.

Following the conventions of \arplsha, the scalar $\Phi$ at the special point with $r=\tilde n_c$, splits into an adjoint field $\phi$ of the $SO(r)$ group and $(n_c-r)/2$ scalars $\{\psi_k\}$ belonging to
the $U(1)$ $N=2$ multiplets.  The following is the effective superpotential governing the light degrees of freedom at the special point after addition of a microscopic adjoint mass term:
\eqn\apssup{W=\sqrt 2 Q_a^i\phi_{ab}Q_b^j\bJ_{ij} +\sqrt 2\sum_{k=1}^{n_c-n_f+2}\psi_ke_k\tilde{e}_k
+\mu \left( \Lambda\sum_{i=1}^{n_c-n_f+2}x_i\psi_i+{1\over 2}\Tr{\phi^2}  \right).}
The $x_i$ are simply dimensionless numbers of order 1 while the $Q'$s are the $n_f$ light hypermultiplets that appear at the root of the $r$-Higgs branch.
These are to be thought of as analogues of the dual quarks present in Seiberg's description of certain $N=1$ gauge theories.
The quark mass perturbation shows up in the low-energy superpotential as a correction,
\eqn\masspert{\Delta W={1\over 2}{m_{ij}}Q^i_a
Q_a^{j}-S_k^im_i\tilde{e}_ke_k.}
Here $S_k^i$ represents the $i-$th quark charge of the monopole
multiplet $e_k$.  It appears in the central extension of the
supersymmetry algebra and consequently contributes to the masses of
the BPS monopoles \swtwo.  Since the theory at the $r$-branch root is
IR free, the degrees of freedom appearing in \apssup\ are indeed
weakly interacting and all the vacua of the resulting theory should be
accessible via a straightforward semiclassical analysis of the
equations of motion of the effective theory.  The solutions of the F
and D-term equations
\eqn\apsmon{ e_k = {\tilde e}_k \sim \sqrt{\mu \Lambda},}
\eqn\apspsi{ ( \sqrt 2 \psi_k - \sum_{i} S_k^i m_i) e_k=0,}
\eqn\apsdphi{ [ \phi, \phi^{\dagger}]=0,}
\eqn\apsdq{{\rm Im} \, Q_a^{i \dagger} Q_b^i =0,}
\eqn\apsfphi{ \sqrt2 \, Q_a^i Q_b^j {\bJ}_{ij} + 2 \, \mu \, \phi_{ab}
=0,}
\eqn\apsfq{ \sqrt2 \, \phi_{ab} Q_b^j {\bJ}_{ij} + m_{ij} Q_a^j=0.}
are similar to the classical solutions of the microscopic theory.  By
performing appropriate gauge rotations $\phi$ can be put in the form
\eqn\apsphi{ \phi= \pmatrix{ \pmatrix{ &\phi_1 \cr -\phi_1 &} & & &
\cr
& \pmatrix{ &\phi_2 \cr -\phi_2 &} &  & \cr  & & \ddots  & \cr
& & & \pmatrix{ &\phi_{[{\tilde n}_c/2]} \cr -\phi_{[{\tilde n}_c/2]}
&}}.}
For an odd ${\tilde n}_c$ there is a null ${\tilde n}_c$-th column and
null ${\tilde n}_c$-th row above.
The solutions can be again classified according to the number of the
nontrivial pairs of ``eigenvectors'' $Q_a^i$ with eigenvalues $\pm
m_i / \sqrt{2}$.  A solution with $r$ nonzero eigenvalues has
$\{\phi_i\}=\{m_i\}$ and
\eqn\apsone{ Q^i_a= \pmatrix{ d_1 & 0 & 0 &\cdots &0 \cr
-id_1 & 0 & 0 & \cdots & 0 \cr
0 & d_2 &0 &\cdots & 0&0\ldots\cr
0 & -id_2& 0 &\cdots & 0&0\ldots\cr
\vdots& \vdots & \vdots &\vdots &\vdots&0\ldots \cr
0 & 0 & 0 & d_r &0&0\ldots\cr
0 & 0 & 0 & -id_r &0&0\ldots\cr
\vdots&\vdots&\vdots&\vdots&\vdots&\vdots\cr}.}
\eqn\range{{\rm where} \qquad 0 \leq r\leq \left[ {\tilde n_c\over 2} 
\right] \, .}
These characterize the VEVs of the squark flavors $i=1,\ldots,n_f$,
while the squark condensates with $i=n_f+1, \ldots , 2n_f$ are of the form
\apsone\ with $d_r\rightarrow\tilde d_r$ and $-id_r \rightarrow
i\tilde d_r$, and with\foot{The possible overall minus sign of $d_{i}$'s can be rotated away by a center element of $SU(n_{f}) \subset USp(2n_{f})$.}
\eqn\apsvev{|d_r|=|\tilde d_r|;\;\; {\rm and}\;\; {\rm Re} (d_r\tilde{d}_r)=\sqrt{ - \mu m_r \over 2}.}

Although these solutions are superficially similar to the classical
solutions of the microscopic theory, the physics is clearly very
different.  The theory at the $r-$branch root is IR free
($n_f>\tilde{n}_c$) and so the values of $r$ can run only from $r=0$
to $r=[\tilde n_c/2]$.

The solution with $r$ pairs of nonvanishing elements of $\phi$ leaves
$SO({\tilde n}_{c}-2r) $ gauge symmetry unbroken with $\tilde
n_c-2r-2$ vacua.  Therefore, summing over all such vacua there are
\eqn\apsntwo{ {\cal N}_2 = \sum_{r}^{[{\tilde n}_c/2] } w({\tilde
n_c}-2r) \, {}_{n_f}\!C_r + {}_{n_f}\!C_{{\tilde n}_c/2}}
ground states with all VEVS (transforming nontrivially under flavor)
vanishing in the $m_i\to 0$ limit.

It is clear from \apsntwo\ that not all the vacua of the $N=1$ theory
arise from perturbing the theory at the special point.  The latter
seems to yield only the vacua with vanishing VEVs i.e. the
$\USp(2n_f)$-symmetric ones.  Therefore the APS effective theory
appears to miss the complete vacuum structure of the softly broken
$N=2$ theory.  We must therefore understand better the nature of and
the role played by the other group of vacua which cannot be associated
with the theory at the special point.

To understand the microscopic origin of the remaining ${\cal N}_1$
vacuum states we turn to the hyperelliptic curves describing the
Coulomb branch of the corresponding $N=2$ theory.  The special point
is associated with a maximal degeneration of the curve and corresponds
to mutually local, charged monopoles becoming light.  However, the
degeneration of the Seiberg-Witten curve can also occur in a way that
describes mutually non-local particles becoming simultaneously
massless leading to an interacting $N=2$ superconformal field theory
\argdoug\scftone.

As we demonstrate in the next section, in the massless ($m_i=0$)
theory there are additional maximal singularities on the ${\cal N}=2$
moduli space where the low-energy effective theories are critical
(scale-invariant).  We refer to these singularities as the ``Chebyshev
points''.  The Chebyshev points split up under a generic mass
perturbation and subsequently yield the first group of ${\cal N}_1$
vacua when an adjoint mass is added.  The enumeration of vacua and
their symmetry-breaking patterns can be best understood in the case
where the quark masses are equal and non-zero -- $m_i=m_0\neq 0$.

The equal mass theory has singularities where the curve has the form
\eqn\critcurve{\eqalign{y^2 & \propto(x-m_0^2)^{2r};\;\;0\leq r\leq
[n_f/2],\cr y^2 & \propto(x-m_0^2)^{2(n_f-r)};\;\;n_f\geq r>[n_f/2].}}
Since the underlying flavor group of the equal-mass theory is
$U(n_f)$, using the results of \eguhorione\ and \eguhoritwo\ we know
that the theories at such singular points are in the universality
class of $SU(r)$ (or $SU(n_f-r)$) gauge theories with $n_f$ flavors. 
Depending on the values of $r$ these are IR-free gauge theories (Class
1 according to \eguhorione) or non-trivial SCFTs when $r=n_f/2$ (Class
3 theories).  Except the cases with $r=n_f/2$ they can be described by
the effective Lagrangeans at the $r-$branch roots of $SU(n_c)$, $N=2$
theories, derived by Argyres, Plesser and Seiberg \arplsei.  The
vacuum structure of the $r$-branch theories was studied in detail in
\ckm\ and it was shown that the theory at $r$-branch root yields
${}_{n_f}C_r$ vacua with flavor symmetry $U(r)\times U(n_f-r).$ Taking
discrete symmetry factors into account, the total number of vacua from
the Chebyshev points is then \eqn\chebyvacua{{\cal
N}_1=(n_c-n_f-2)2^{n_f},} which correctly reproduces the number of
vacua with finite VEVs.

When the masses are sent to zero the $r$-branch roots merge, the
criticality of the hyperelleptic curve changes and the flavor symmetry
of the microscopic theory is enlarged to $\USp(2n_f)$.  We believe
that this point (the Chebyshev point) describes an SCFT belonging to a
new universality class and which is strongly-interacting due to the
appearance of light, mutually non-local degrees of freedom.

%: Remarks on suggested derivations of $N=1$-duality
\subsec{Remarks on suggested derivations of $N=1$-duality}

At this point we would like to make a few remarks about the possible derivation of $N=1$ Seiberg duality which was suggested by Argyres, Plesser and Shapere in \arplsha.
This ``derivation'' was made possible by the idea that by moving along the $N=2$ moduli space and by smoothly changing $\mu$, we can continuously interpolate between the small-$\mu$ description of the special point $SO(\tilde n_c)$ theory to the large-$\mu$ $SO(n_c)$ gauge theory with $N=1$
supersymmetry.
However, we have seen above that the $N=1$ theory obtained by introducing an adjoint mass has vacua emerging both from the special point and the superconformal Chebyshev point (to be
studied below).
The question then is, what role (if any) do the Chebyshev vacua play in the derivation of
$N=1$ duality?
To understand this it is instructive to look at the behaviour of the condensate $u_2 = \langle \Tr \Phi^2 \rangle$ in the $N=1$ limit at these two points, namely the Chebyshev point and the special point.
Upon integrating out the adjoint field $\Phi$, we have an $N=1$ theory with dynamical scale $\Lambda_{n_c, 2n_f}^{3n_c - 2n_f -6} = \mu^{n_c-2} \Lambda_{N=2}^{2n_c - 2n_f-4}$.
As seen in the previous section, the low energy theory is the Intriligator-Seiberg magnetic theory with dual quarks and mesons, with a superpotential $W={1 \over 2 \kappa} qMq - {1 \over 2 \mu} \Tr M \bJ M \bJ$.

At the special point, which is $\USp(2n_f)$ symmetric, neither $M$ nor $q$ acquire expectation values and hence $W=0$. This automatically implies that $u_2=0$ at the special point, since it is the $\mu$ derivative of the expectation value of the superpotential. 
On the other hand, at the Chebyshev point, where $\det M \neq 0$, we can integrate out all the magnetic quarks to obtain a pure Yang-Mills theory at low energy with a gaugino condensate-induced superpotential for the mesons \dualsup.  
As was discussed in detail in the previous section, by minimizing the superpotential we find $\langle M \rangle \sim \mu \Lambda_{N=2}$. 
This implies that $W \sim \mu \Lambda_{N=2}^2$, and consequently $u_2 \sim \Lambda_{N=2}^2$ at the Chebyshev point.
The ``distance'' between the special and the Chebyshev points at fixed $\Lambda_{N=2}$ remains the same in the large $\mu$ limit. However, in the true $N=1$ limit wherein $\Lambda_{n_c, 2n_f}$ is kept fixed, the locations of the two singularities appear to merge (since $\Lambda_{N=2} \rightarrow 0$). It must be pointed out that the vanishing of $u_2$ alone does not imply merging of the two points. One should look at the expectation values of all the other gauge invariant order parameters.

This suggests that in order to understand the physics of $N=1\, $ $ SO(n_c)$ gauge theory with $2n_f>n_c-2$ flavors, one must understand the physics of both sets of vacua.
Consequently, from the viewpoint of the corresponding $N=2$ theory it appears that we must investigate the descriptions of $both$ the special point and the Chebyshev point in order to understand the origin of Seiberg duality in the $N=1$ limit.  
We therefore see that the derivation of Seiberg duality as presented in \arplsha\ was incomplete in this aspect.

%: Quark mass perturbation of hyperelleptic curves
\newsec{Quark mass perturbation of hyperelleptic curves}

We will now investigate the low-energy physics on the Coulomb branch of the $N=2$ gauge theory $(\mu=0)$ which is described as usual by the moduli space of certain hyperelliptic curves.  The complex structure $\tau$ of the curves controls the physics of the Coulomb branch in that it is identified with the effective gauge-coupling of the low energy theory.  Singularities on the Coulomb branch due to the appearance of new massless states in the low-energy theory are signalled by the degeneration of the hyperelliptic curves -- where certain specific cycles of the Riemann surface associated with the curve pinch off.  Some of these singularities are rather special. 
These are the Coulomb branch singularities which, after $N=2$-breaking perturbations $(\mu\neq0)$ become vacua of the resulting theory with $N=1$ supersymmetry.  Therefore, by inspecting the hyperelliptic curves of the $N=2$ $SO(n_c)$ gauge theory with $n_f$ hypermultiplets we can identify which points on the Coulomb branch correspond to $N=1$
vacua and how many such points there are.  Furthermore we will also be able to identify the two groups of vacua and unravel the physics of the first group of vacua whose origin {\it cannot} be explained by the effective picture of the previous section {\it \`a la} Argyres-Plesser-Shapere \arplsha.

Our first task will be to identify the points on the Coulomb branch of the $N=2$ theory which give rise to the first group ${\cal N}_1$ of $N=1$ vacua and, as is evident from the discussion in the previous section, these are distinct from the so-called special point of APS.
Since they survive the breaking to $N=1$, they must be points of maximal degeneracy i.e. where $[{n_c\over 2}]$ pairs of branch points coincide.  It is well-known that such singularities may be of two distinct types: a) Where the light degrees of freedom are mutually local, as in the case of the special point at the $r-$branch roots; b) where mutually non-local particles are simultaneously massless.  Theories of the second type with $N=2$ supersymmetry were argued to be superconformal in \scftone.  We will argue below that the first group of $N=1$ vacua emerge upon perturbing such superconformal theories.

The curve for the $SO(n_c)$ $N=2$ theory is \argshap,
\eqn\soncurve{ y^2 = x\prod_{a=1}^{[n_c/2]} (x- \phi_a^2)^2 - 4
\Lambda^{2(n_c-2-n_f) } x^{2+\epsilon} \prod_{i=1}^{n_f} (x-m_i^2), }
where $\epsilon =1$ if $n_c$ is even and $\epsilon =0$ if $n_c$ is
odd.  Since we are mainly interested in the limit where the bare
masses $m_i$ of the quark hypermultiplets are vanishingly small, we
first consider the curves with $m_i=0$,
\eqn\nomass{ y^2 = x\prod_{a=1}^{n_c/2} (x- \phi_a^2)^2 - 4
\Lambda^{2(n_c-2-n_f) } x^{n_f+3}, \quad n_c=2 \ell,}
\eqn\zeromass{ y^2 = x\prod_{a=1}^{(n_c-1)/2} (x- \phi_a^2)^2 - 4
\Lambda^{2(n_c-2-n_f) } x^{n_f+2}, \quad n_c=2\ell +1.}
We will subsequently treat the effects of the non-zero masses as a
perturbation and then enumerate the $N=1$ vacua and their properties.

As before we will treat the equal mass case $m_i=m_0\neq 0$ and the
case of generic unequal masses separately.  For either of these two
situations, the actual form of the curves depends on whether $n_f$ and
$n_c$ are even/odd and hence we must consider four independent
possibilities.

%: Unequal mass perturbation at Chebyshev points
\subsec{\bf Unequal mass perturbation at Chebyshev points}

\item{i)} {\bf $n_f$ even, $n_c$ even }

The singular points which yield the first group of vacua can be
located (when $m_i=0$) by following the methods of \brandland\ and
\dougshenk.  This is achieved by first choosing $n_f/2 +1 $ of
$\phi_a$'s to be vanishing so that the curve \nomass\ has the form
\eqn\evencurve{ y^2 = x^{n_f+3} \left[\prod_{a=1}^{(n_c- n_f -2)/2}
(x- \phi_a^2)^2 - 4 \Lambda^{2(n_c-2-n_f) } \right] .}
The remaining $\phi_a$'s can be obtained by using the properties of
Chebyshev polynomials.  The Chebyshev polynomial $T_N(x)$ of order $N$
is defined as
\eqn\chebydef{ T_N (x) = \cos (N \arccos x) = 2^{N-1} \prod_{k=1}^N
(x-w_k),}
with $w_k= \cos \pi (k-{1\over 2})/N,$ $k=1,2,\ldots N$.  Note that
for $N$ odd, there exists a value of $k$, $k=(N+1)/2$, such that
$w_k=0$.  Note also that $w_{N-k+1} = -w_k$, so that for even $N$
\eqn\chebyeven{ T_N (\sqrt x /2 ) = { 1\over2} \prod_{k=1}^{N/2} (x- 4
w_k^2) \, ;}
while for odd $N$
\eqn\chebyodd{ T_N (\sqrt x /2 ) = { 1\over 2} \sqrt x
\prod_{k=1}^{(N-1)/2} (x - 4 w_k^2) \, ,}
These properties will be useful in what follows.

Now we identify the first term in the brackets in Eq.~\evencurve\ with
$2T_{n_c-n_f-2}(\sqrt x/2)$ and set $\Lambda=1$ for notational
convenience.  Then we must have $\phi_a^2 = 4 w_a^2,$ with $w_a= \cos
\pi (a-{1\over 2})/N,$ $a=1,2,\ldots N/2$, and $N=n_c- n_f -2$.  At
this point, which we will henceforth refer to as the ``Chebyshev
point'', the curve becomes
\eqn\unitlambda{ y^2 = 4 \, x^{n_f+3} \left[ T_{n_c- n_f -2}(\sqrt{x}/2)^2
- 1 \right] \equiv - 4 x^{n_f+3} \sin^2 \left[ (n_c- n_f -2) \arccos {\sqrt x\over
2} \right].}
This curve has a zero at the origin of order $n_f + 4$, and $N/2-1 =
(n_c-n_f-2)/2 - 1$ double zeroes at $x= 4 \cos^2 {k\pi /N}$, $ k=1,2,
\ldots (N/2-1)$ and a single zero at $x=4$.  Thus we have located a
point of maximal degeneracy.

Upon a generic mass perturbation, i.e.~introduction of
$N=2$--preserving quark masses $m_i \ll \Lambda_{N=2}$, the point of
maximal degeneracy will not only shift somewhat but will also split
into a {\it group} of singularities.  Our goal is to find the nature
and number of points in this singular group.

We will first consider the special case where $n_f=n_c-4$.  The more
general problem can be reduced to this example, as we will argue
subsequently.  At the Chebyshev point, the curve for this theory is
\eqn\spcl{ y^2 = x^{n_f+3} [(x- \phi^2)^2 - 4 \Lambda^4]= x^{n_f+3} [x- \phi^2 +2 \Lambda^2] [x- \phi^2 - 2 \Lambda^2].}
and $\phi^2= \pm 2 \Lambda^2$ (this multiplicity corresponds to the
discrete $\bZ_{n_c-n_f-2}$ symmetry factor with $n_c-n_f -2 =2$).  The
zero at the origin is of order $n_f +4$, and there is a single zero at
$4 \Lambda^2$ (or $- 4 \Lambda^2$).  Let us consider the point
$\phi^2= 2 \Lambda^2$.

Denoting the positions of the new singular points in the $N=2$ moduli
space as
\eqn\pert{ \phi_a^2= (\phi_1^2, \phi_2^2, \ldots, \phi_{n_c/2-1}^2, 2
\Lambda^2 + \delta \phi^2),}
the perturbed curve
\eqn\identicalone{ y^2 = x\prod_{a=1}^{n_c/2 -1} (x- \phi_a^2)^2 (x- 2
\Lambda^2 - \delta \phi^2)^2 - 4 \Lambda^{4} x^{3} \prod_{i=1}^{n_f}
(x-m_i^2),}
must be identical to
\eqn\identicaltwo{ y^2 = \prod_{a=1}^{n_c/2} (x - \alpha_a)^2 (x -
\beta).}
Here, the unperturbed zero of order $n_f+4$ has been assumed to split
into $n_f/2+2$ double zeros at $x=\alpha_a$ some of which could be
degenerate.  The important point to note is that the $\alpha_a$ are
necessarily ``small'' and proportional to positive powers of the $m_i$
so that they vanish in the zero mass limit.  On the other hand $\beta$
is the perturbed value of the single zero at $4\Lambda^2$.

The explicit factor of $x$ in Eq.~\identicalone\ means that one of
the $\alpha_a$'s is zero.  However, this in turn implies that
\identicalone\ has actually at least a double zero at the origin. 
Since the second term of \identicalone\ has a cubic zero there, it
means that one of the $\phi_a$'s is zero:
\eqn\pert{ \phi_{n_c/2-1}^2 = 0}
and consequently, \identicaltwo\ has at least a cubic zero, so that in
fact {\it two } of the $\alpha_a$'s are zero.  To obtain the location of the critical points we must then solve the following identity ($y^2 = x^3 F(x))$:
\eqn\identity{\eqalign{ F(x)=&\prod_{a=1}^{n_c/2 -2} (x- \phi_a^2)^2 (x- 2 \Lambda^2 - \delta \phi^2)^2 - 4 \Lambda^{4} \prod_{i=1}^{n_c-4} (x-m_i^2) \cr =& \;\;x \prod_{a=1}^{n_c/2 -2} (x- \alpha_a)^2 (x - \beta)}.}

Using $F(0)=0$ one can argue that
\eqn\shiftorder{\phi_a^2\sim m_i^2\, ; \qquad \delta\phi^2\sim m\Lambda,}
while comparison of the coefficient of $x$ on both sides reveals that all $\alpha_a$ except one are of order $m^2$:
\eqn\orderalpha{\alpha_1,\alpha_2,\ldots\alpha_{n_c/2-3}\sim m^2 \; ;
\qquad \alpha_{n_c/2-2}\sim m\Lambda.}
Similarly one may also show that
\eqn\deltaphi{\delta\phi^2\approx2\alpha_{n_c/2-2} \, .}
In the $x$-plane, we must then have for $|x|\sim m^2$ (recall $y^2 = x^3 F(x)$),
\eqn\id{ F(x)= \prod_{a=1}^{n_c/2 -2} (x- \phi_a^2)^2 - \prod_{i=1}^{n_c-4}
(x-m_i^2)= c \, x \prod_{b=1}^{n_c/2-3} (x- \alpha_b)^2,}
where $c\approx4\Lambda^2\alpha_{n_c/2-2}^2$.  Moving terms around, we
get
\eqn\moving{\eqalign{& \left[ \prod_{a=1}^{n_c/2 -2} (x- \phi_a^2) +\sqrt
{cx} \prod_{b=1}^{n_c/2-3} (x- \alpha_b) \right] \cdot \left[ \prod_{a=1}^{n_c/2
-2} (x- \phi_a^2) - \sqrt {cx} \prod_{b=1}^{n_c/2-3} (x- \alpha_b) 
\right]
\cr
&= \prod_{i=1}^{n_c-4} (x-m_i^2),}} which, under the variable change
$x=z^2$ leads to 
\eqn\sopra{\eqalign{& \left[ \prod_{a=1}^{n_c/2 -2} (z^2- \phi_a^2)
+\sqrt{c} \, z\prod_{b=1}^{n_c/2-3} (z^2- \alpha_b) \right] \cdot
\left[ \prod_{a=1}^{n_c/2 -2} (z^2- \phi_a^2) - \sqrt{c} \,
z\prod_{b=1}^{n_c/2-3} (z^2- \alpha_b) \right] \cr =&
\prod_{i=1}^{n_c-4} (z + m_i) (z-m_i).}}
Now note that the operation $z\to -z$ exchanges the two terms on the
left hand side as well as the factors on the right hand side.  We also
note that the masses $m_i$ are all generic and distinct.  We may
therefore equate each factor on the left to a product of $n_c-4$ {\it
distinct} $(z\pm m_i)$ factors so that the resulting equations get
exchanged under $z\to -z$.  For instance, one possible set of
solutions may be obtained by solving
\eqn\allsign{\prod_{a=1}^{n_c/2 -2} (z^2- \phi_a^2) +\sqrt{c} \, z
\prod_{b=1}^{n_c/2-3} (z^2- \alpha_b)=\prod_{i=1}^{n_c-4} (m_i +z).}
Dividing the expression on the right hand side into even and odd
powers of $z$, we note that the sum of the even terms is a polynomial
in $z^2$ of order ${n_c/2 -2}$ with coefficients which are functions
of $m_i^2$.  The even terms must therefore be identified with
$\prod_{a=1}^{n_c/2 -2} (z^2- \phi_a^2)$.  Consequently, the set
$\{\phi_a^2\} $ is determined uniquely (up to Weyl group
transformations) in terms of quark masses.  Similarly, identifying the
odd terms from $\prod_{i=1}^{n_c-4} (m_i +z)$, with $\sqrt{c} \, z
\prod_{b=1}^{n_c/2-3} (z^2- \alpha_b)$ uniquely determines $\sqrt c$
and the set $\{\alpha_b\}$.  The location of the singularities on the
Coulomb branch is determined uniquely by the set $\{\phi_a^2\}$ and
$\sqrt c$, the latter being directly proportional to $\delta\phi^2$.

All other possible solutions are found by choosing different signs in
front of $z$ in Eq.~\allsign.  (As long as all masses are different,
the reduction from Eq.~\sopra\ to Eq.~\allsign\ is consistent). 
Since there are $2^{n_c-4}=2^{n_f}$ possible distinct sign choices we
find $2^{n_f}$ solutions.  The copies generated by the action of the
$\bZ_{n_c-n_f-2}$ discrete symmetry factor yield a total of ${\cal
N}_1=(n_c-n_f-2)2^{n_f}$ points of maximal degeneracy each of which
becomes an $N=1$ supersymmetric vacuum upon introducing an adjoint
mass.

Finally, let us now see how the general $n_f$ even, $n_c$ even cases
reduce to the problem studied above.  This can be easily deduced from
the form of the hyperelleptic curve near the Chebyshev point of the
$m_i=0$ theory.  Writing the curve as
\eqn\general{\eqalign{y^2=&x\prod_{a=1}^{n_f/2+1}(x-\phi_a^2)^2
\prod_{k=1}^{(n_c-n_f-2)/2}(x-\phi_k^2)^2 -4x^3\Lambda^{2(n_c-n_f-2)}
\prod_{i=1}^{n_f}(x-m_i^2),\cr
=&\prod_{a=1}^{n_c/2}(x-\alpha_a)^2(x-4\Lambda^2-\beta),}}
we note that near the Chebyshev point (the maximally degenerate point)
$n_f/2+1$ $\phi_a$'s must be ``small'' (i.e. vanishing in the zero
mass limit) while the rest, being (perturbed) roots of the Chebyshev
polynomial of the $m_i=0$ theory must all be finite and proportional
to the $N=2$ dynamical scale.  The fact that the $m_i=0$ curve in
Eq.~\unitlambda\ has a zero at the origin of order $n_f+4$,
automatically implies that only $n_f/2+2$ of the $\alpha_a$ are
vanishingly small, the rest being determined essentially by the $N=2$
dynamical scale.  We further know from arguments following
Eqs.~\identicalone\ and \identicaltwo\ (which apply in the general
case as well), that one of the $\phi_a$'s and two of the $\alpha_a$
are identically zero.  Putting all these requirements together, for
$|x|\ll\Lambda_{N=2}$, the curve for the theory has the form
\eqn\gencurve{ \eqalign{ y^2 \approx &  4 \Lambda^{2(n_c-  n_f-2)/2}  \; x^3 \left[  \prod_{a=1}^{n_f/2}(x-\phi_a^2)^2 - \prod_{i=1}^{n_f}(x-m_i^2)  \right]  \cr & =4 \Lambda^{2(n_c-  n_f-2)/2}  \; x^4
\;\prod_{a=1}^{(n_f/2-1)}(x-\alpha_a)^2 \, . }}
Note that this is the form of the curve in Eq.~\id\ and consequently the arguments determining the number of solutions to this identity are identical.
The only difference is in the number of copies resulting from the discrete symmetry factor $(n_c-n_f-2)$.  Therefore the total number of $N=1$ vacua from the Chebyshev points is ${\cal N}_1 = 2^{n_f}(n_c-n_f-2)$.

\item{ii)}{\bf $n_f$ even, $n_c$ odd}

The Chebyshev point is once again characterized by $n_f/2 +1$
vanishing $\phi_a$'s.  Setting $\Lambda=1$ for notational convenience,
the curve becomes
\eqn\zeromass{ y^2 = x^{n_f+2} \left[ x \prod_{a=1}^{(n_c- n_f -3)/2}
(x- \phi_a^2)^2 - 4 \right].  }
Identifying the first term in the brackets with $ (2 T_{n_c- n_f
-2}(\sqrt x /2 ))^2 $ the curve for the massless theory at the
Chebyshev point is
\eqn\cheb{ y^2 = 4 x^{n_f+2} \left[ T_{n_c- n_f -2}(\sqrt{x}/2)^2 - 1
\right]= - 4 x^{n_f+2} \sin^2 \left[ (n_c- n_f -2) \arccos {\sqrt
x\over 2} \right].}
Now the order of the zero at the origin is $n_f+2$; there is a single
zero at $x=4$ and $(N-1)/2=(n_c-n_f-3)/2$ double zeroes at
$x=4\cos^2(k\pi/N);\; k=1,2,\ldots,(N-1)/2$.  As before, it is
sufficient to restrict our analysis to a special case with $n_f =
n_c-3$, the more general analysis being similar.  For $n_f=n_c-3$ the
curve at the superconformal point is
\eqn\spclSH{ y^2 = x^{n_f+2} (x- 4 \Lambda^2 ).}
All $\phi_a$'s are zero, and this is consistent with the fact that the
discrete symmetry factor is $n_c-n_f-2=1$.  Upon perturbation then,
all the $\phi_a$'s get small non-zero VEVs,
\eqn\perturb{ \phi_a^2= (\phi_1^2, \phi_2^2, \ldots,
\phi_{(n_c-1)/2}^2),}
($\phi^2$ all small) and the curve
\eqn\curve{ y^2 = x\prod_{a=1}^{(n_c-1)/2} (x- \phi_a^2)^2 - 4
\Lambda^{2 } x^{2}\prod_{i=1}^{n_f} (x-m_i^2),}
must be identical to
\eqn\identicaltt{y^2 = \prod_{a=1}^{(n_c-1)/2} (x - \alpha_a)^2 (x - 4
\Lambda^2 +\beta).}
This identity requires one of the $\phi_a$ and one of the $\alpha_a$
to vanish identically:
\eqn\vanish{ \phi_{(n_c-1)/2}=0; \quad \alpha_{(n_c-1)/2}=0,}
leading to the condition,
\eqn\newident{ x\prod_{a=1}^{(n_c-3)/2 } (x- \phi^2)^2 - 4 \Lambda^{2
} \prod_{i=1}^{n_c-3} (x-m_i^2) = \prod_{a=1}^{(n_c-3)/2} (x -
\alpha_a)^2 (x - 4 \Lambda^2 +\beta).}
This is solved by assuming
\eqn\solve{ \phi_1^2, \ldots, \phi_{(n_c-5)/2 }^2 \sim m^2, \quad
\phi_{(n_c-3)/2}^2 \sim m \Lambda, \quad \alpha_a \sim m^2.}
In fact, near $x \sim m^2$, the curve can be rewritten in the
following form after the variable change, $x=z^2$,
\eqn\set{\eqalign{ & \left[ \prod_{a=1}^{(n_c-3)/2} (z^2 - \alpha_a) +
\sqrt{c} z\prod_{a=1}^{(n_c-5)/2 } (z^2- \phi^2) \right] \cdot \left[
\prod_{a=1}^{(n_c-3)/2} (z^2- \alpha_a)\right.\cr 
&\left.-\sqrt{c}z\prod_{a=1}^{(n_c-5)/2 } (z^2- \phi^2) \right] 
=\prod_{i=1}^{n_c-3} (m_i +z)(m_i -z).}}
Here $\sqrt c\approx \phi_{(n_c-3)/2}^2/2\Lambda$.  The rest of the
arguments are identical to the previous case and we thus find
$2^{n_c-3}=2^{n_f}$ solutions.  We must of course include the discrete
symmetry factor as well so that we find ${\cal N}_1 =(n_c-n_f -2)
2^{n_f}$ vacua.  Similar arguments apply for the general case with
$n_f$ even and $n_c$ odd.

\item{iii)} {\bf $n_f$ odd,  $n_c$ even}

The Chebyshev point is characterized by $(n_f+3)/2$ vanishing
$\phi_a$'s and the curve (taking $\Lambda=1$):
\eqn\trial{y^2 = x^{n_f+3} \left[ x\prod_{a=1}^{(n_c- n_f -3)/2} (x-
\phi_a^2)^2 - 4 \right].  }
The first term in parentheses can be identified with $ (2 T_{n_c- n_f
-3}(\sqrt x /2 ))^2 $ so that the curve can be rewritten as
\eqn\rewritten{ y^2 = 4 x^{n_f+3} [ T_{n_c- n_f -2}(\sqrt{x}/2)^2 -
1]= - 4 x^{n_f+3} \sin^2 \left[ (n_c- n_f -2) \arccos {\sqrt x\over 2} 
\right] \; ,}
with a zero at the origin of order $n_f+3$.  Focussing attention on
the special case with $n_f = n_c-3$, the curve takes the form
\eqn\form{ y^2 = x^{n_f+3} (x- 4 \Lambda^2 ).}
All  $\phi_a$'s are zero, again consistent with the discrete symmetry
factor being $n_c-n_f-2=1$. Upon perturbation the small VEVs
$\phi_a^2= (\phi_1^2, \phi_2^2, \ldots, \phi_{n_c/2}^2),$
modify the curve to
\eqn\curveone{ y^2 = x\prod_{a=1}^{n_c/2} (x- \phi_a^2)^2  - 4 
\Lambda^{2 }
x^{3}\prod_{i=1}^{n_f} (x-m_i^2),}
which must be identical to
\eqn\identit {y^2 =   \prod_{a=1}^{n_c/2} (x - \alpha_a)^2  (x - 4
\Lambda^2  +\beta).}
Now, we must require {\it two} $\alpha_a$'s and one $\phi_a$ to
vanish identically,
\eqn\vanishing{\phi_{n_c/2} =0;      \quad  
\alpha_{n_c/2}=\alpha_{n_c/2-1}=0,}
and we find an identity of the form
\eqn\find{ \prod_{a=1}^{n_c/2-1} (x- \phi_a^2)^2  - 4 \Lambda^{2 }
\prod_{i=1}^{n_c-3} (x-m_i^2)
=    x\prod_{a=1}^{n_c/2-2} (x - \alpha_a)^2  (x - 4 \Lambda^2  
+\beta),}
which can be solved consistently by assuming
\eqn\sol{    \phi_1^2, \ldots, \phi_{n_c/2 -2}^2  \sim  m^2, \quad
\phi_{n_c/2-1}^2 \sim m \Lambda, \quad
\alpha_a \sim m^2.}
As in the earlier cases, for small $x$ $ \sim m^2$, the identity
(after the customary variable change) assumes the form:
\eqn\changevar{\eqalign{& \left[ z\prod_{a=1}^{n_c/2-2} (z^2 -
\alpha_a) + \sqrt{c} \prod_{a=1}^{n_c/2-2} (z^2- \phi_a^2) \right]
\cdot \left[ z\prod_{a=1}^{n_c/2-2} (z^2- \alpha_a) - \sqrt{c}
\prod_{a=1}^{n_c/2-2} (z^2- \phi^2) \right]\cr &= \prod_{i=1}^{n_c-3}
(z+m_i)(z-m_i),}}
where $\sqrt c\approx\phi^2_{n_c/2-1}/2\Lambda$.
This has $2^{n_f}$ distinct solutions with $n_c-n_f-2$ copies. 
Extending the above arguments to the general case, we find that the
Chebyshev (superconformal) point yields ${\cal N}_1 =(n_c-n_f-2)
2^{n_f}$ vacua upon mass perturbation.

\item{iv)}{\bf $n_f$ odd,  $n_c$ odd}

We now turn to the final possibility wherein both $n_f$ and $n_c$ are
odd.  The Chebyshev solution is obtained by taking $(n_f+1)/2$ of the
$\phi_a$'s to be zero:
\eqn\curveagain{ y^2 = x^{n_f+2} \left[ \prod_{a=1}^{(n_c- n_f -2)/2} (x-
\phi_a^2)^2 - 4 \right] \, .}
The first term in parentheses can be identified with the square of the
Chebyshev polynomial $T_{n_c-n_f -2}(\sqrt x/2 )$ and the curve
becomes:
\eqn\redo{ y^2 = 4 x^{n_f+2} [ T_{n_c- n_f -2}(\sqrt{x}/2)^2 - 1]= - 4
x^{n_f+2} \sin^2 \left[ (n_c- n_f -2) \arccos {\sqrt x\over 2} \right].}
Now the order of the zero at the origin is $n_f+3$.  For the special
case with $n_f = n_c-4$, the curve is of the form
\eqn\curveform{ y^2 = x^{n_f+2} ((x-\phi^2)^2- 4 \Lambda^4 ).}
All except one of the $\phi_a$'s are zero and the non-zero $\phi_a$
can have two possible values $\pm 2\Lambda^2$ which is consistent with
the $\bZ_2$ discrete symmetry.  Upon perturbation, $\phi_a^2=
(\phi_1^2, \phi_2^2, \ldots, \phi_{(n_c-1)/2}^2)$, where all the VEVs
are small except $\phi_{(n_c-1)/2}^2\sim \Lambda^2$, the curve takes
the form
\eqn\curvetwo{ y^2 = x\prod_{a=1}^{(n_c-1)/2} (x- \phi_a^2)^2 - 4
\Lambda^{4 } x^{2}\prod_{i=1}^{n_f} (x-m_i^2),}
which must be identical to
\eqn\identitbis{ y^2 = \prod_{a=1}^{(n_c-1)/2} (x - \alpha_a)^2 (x - 4
\Lambda^2 +\beta).  }
Since the curve \curvetwo\ has a zero at the origin, we conclude that
one of the $\alpha_a$'s is zero which in turn implies that one of the
$\phi_a$'s must be zero also,
\eqn\kieru{\phi_{(n_c-1)/2}=0.}
Therefore, we find
\eqn\findfinal{ x \prod_{a=1}^{(n_c-3)/2} (x- \phi_a^2)^2 - 4
\Lambda^{4} \prod_{i=1}^{n_c-4} (x-m_i^2) = \prod_{a=1}^{(n_c-3)/2} (x
- \alpha_a)^2 (x - 4 \Lambda^2 +\beta).  }
A consistent solution can be found by assuming
\eqn\solution{ \phi_1^2, \ldots, \phi_{(n_c-5)/2 }^2 \sim m^2, \quad
\phi_{(n_c-3)/2}^2=0 \quad
\phi_{(n_c-1)/2}^2=2\Lambda^2+\delta\phi^2.}
Further we also need to assume that
\eqn\moreassump{ \alpha_1, \ldots, \alpha_{(n_c-5)/2)} \sim m^2, \quad
\alpha_{(n_c-3)/2}\sim m\Lambda.}
For small $x \sim m^2$, this allows us to rewrite the above
identity in a simplified form after a variable change $x=z^2$:
\eqn\newvar{\eqalign{& \left[ \sqrt{c}\prod_{a=1}^{(n_c-5)/2} (z^2 -
\alpha_a) +z \prod_{a=1}^{(n_c-5)/2} (z^2- \phi_a^2) \right] \cdot
\left[ \sqrt{c}\prod_{a=1}^{(n_c-5)/2} (z^2- \alpha_a)\right.\cr
&\left.- z
\prod_{a=1}^{(n_c-5)/2} (z^2- \phi^2) \right] 
=\prod_{i=1}^{n_c-4} (z+m_i)(z-m_i).}}
This equation has precisely $2^{n_f}$ solutions with $n_c-n_f-2$ copies
required by the discrete symmetry of the theory.  Generalizing these
arguments appropriately we conclude that for general $n_c$ odd and
$n_f$ odd the total number of $N=1$ vacua generated by mass
perturbation of the Chebyshev point is ${\cal N}_1 =(n_c-n_f-2)
2^{n_f}.$

%: Equal mass perturbation at Chebyshev points
{\subsec{\bf Equal mass perturbation at Chebyshev points}}

We now turn to the case where the quark masses are all equal and
non-zero, $m_i=m_0\neq 0$. Recall that this theory was analyzed
earlier in the large-$\mu$ limit. It has a $U(n_f)$ flavor symmetry
which is spontaneously broken to $U(r)\times U(n_f-r)$. Let us now
see how this comes about in the {\it small}-$\mu$ regime from the
viewpoint of the associated Seiberg-Witten curves.

We study the case where both $n_f$ and $n_c$ are even and $n_f=n_c-4$. 
The generalization of the relevant arguments to different numbers of
flavors and colors is quite straightforward.

When the quark masses are all equal $m_i=m_0\neq 0$, Eq.~\sopra\ takes
the form
\eqn\sopraeq{\eqalign{& \left[ \prod_{a=1}^{n_c/2 -2} (z^2- \phi_a^2)
+\sqrt{c} \, z\prod_{b=1}^{n_c/2-3} (z^2- \alpha_b) \right] \cdot
\left[ \prod_{a=1}^{n_c/2 -2} (z^2- \phi_a^2) - \sqrt{c} \,
z\prod_{b=1}^{n_c/2-3} (z^2- \alpha_b) \right] \cr =&(z^2
-m_0^2)^{(n_c-4)}.}}
A direct application of the arguments of the previous section reveals
that there are only $(n_c-4)(n_c-3)/2$ distinct solutions.  However,
this is not equal to the number of vacua remaining after
SUSY-breaking.  The vacuum number and the physics of these points is
revealed by the form of the curve at these points.  Recall that
solutions to the above equation may be found by equating one of the
factors on the left hand side to the product of any $n_c-4$ terms on
the right hand side so that the two resulting equations are
interchanged under $z\to -z$:
\eqn\solequal{ \prod_{a=1}^{n_c/2 -2} (z^2- \phi_a^2) +\sqrt{c} \;
z\prod_{b=1}^{n_c/2-3} (z^2- \alpha_b)=(z+m_0)^{(n_c-4-r)}(z-m_0)^r.}
Thus solutions may be classified by the integer $r$ ($0\leq r\leq
n_c-4$). When $2r\leq n_c-4$, a factor of $(z^2-m_0^2)^{r}$ appears
and this leads to a solution of Eq.~\solequal\ where $r$ of the
$\phi_a$'s and $r$ of the $\alpha_a$'s are equal to $m_0^2$. The
remaining $\phi_a$ and $\alpha_a$ at this
point can be determined uniquely.

Therefore at this particular singularity the Seiberg-Witten curve
itself has an overall factor of $(z^2-m_0^2)^{2r}=(x-m_0^2)^{2r}$. 
When $2r> n_c-4$, the curve in the vicinity of the singular point is
$y^2\propto (x-m_0^2)^{2(n_c-4-r)}$.  Such degeneracies have been
analyzed and classified in detail in \eguhorione\ and \eguhoritwo.  In
particular in theories with $m_0\neq 0$ and $U(n_f)$ flavor symmetry,
singularities of the above type were shown to be critical points
representing $N=2$ SCFTs.  For $2r\neq (n_c-4)$ these SCFTs are in the
same universality class as the Class 1 theories of Eguchi et al
\eguhorione.  Specifically, when $r<(n_c-4)/2$ the theories at the
critical points are in the universality class of $SU(r)$ gauge
theories with $n_f$ flavors, and that of $SU(n_c-4-r)$ gauge theories
with $n_f$ flavors when $r>(n_c-4)/2$.  These so-called Class 1
theories are trivial (IR free) theories.  The $r=(n_c-4)/2$ theory
flows to a non-trivial fixed point with $SU((n_c-4)/2)$ gauge symmetry
and $n_f$ hypermultiplets.

The emergence of such IR-free and non-trivial superconformal
$SU(r)$ gauge theories was also noticed in $N=2$ $\USp(2n_c)$
theories in \ckm . The physics of the IR-free theories is contained in
the Argyres-Plesser-Seiberg \arplsei\ effective Lagrangeans which
describe the degrees of freedom appearing at the $r$-branch  roots
(non-baryonic roots) in the moduli space of $SU(n_c)$, $N=2$
theories. The $SU(r)$ gauge theories which can be understood
semiclassically survive in the small mass regime as well, due to the
non-renormalization theorems which protect the Higgs branches
emanating from the $r$-branch roots.
Upon perturbation to $N=1$, each $r$-branch root yields
${}_{n_f}C_r$ vacua with global symmetry $U(r)\times U(n_f-r)$ --
this result was also established in \ckm. The flavor symmetry breaking
patterns are identical to those observed semiclassically.
Therefore, including the discrete symmetry factors , the total number
of $N=1$ vacua from the Chebyshev point in the equal mass theory
\eqn\equalnone{{\cal N}_1=(n_c-n_f-2) \left[ \sum_{r=0}^{n_f/2}{}_{n_f}C_r
+\sum_{r=n_f/2+1}^{n_f}{}_{n_f}C_r \right] =(n_c-n_f-2)2^{n_f}}

The above arguments and statements can be directly carried over to all
the other cases, except that when $n_f$ is odd no nontrivial theories
of Class 2 or Class 3 are encountered -- only the IR free theories of
Class 1 are found at the critical points.

What happens if we now send the masses to zero? Clearly, all the
$r-$branch roots merge together in this limit and the curve has the
form $y^2\sim x^{n_c}$ for $n_c$-even and $y^2\sim x^{n_c-1}$ with a
$\USp(2n_f)$ global symmetry. The criticality of the curves and the $\USp(2n_f)$ global symmetry suggests that these theories belong to a new universality class of nontrivial conformal field theories.  We conjecture that these are strongly-interacting theories with mutually non-local degrees of freedom.
Such interacting SCFT's were also encountered in the analysis of the $N=2$, $\USp(2 n_c)$ gauge theories in \ckm. In that context, the monodromies of the singularities were also explicitly checked to  show the appearance of mutually nonlocal light degrees of freedom. 

%: Mass perturbation at the APS special point
\subsec{\bf Mass perturbation at the APS special point}

The analysis of \arplsha\ showed that along the roots of the
$r$-branch there exists a special point with 
$r=\tilde{n}_c=2n_f-n_c+4$ where the
hyperelliptic curve degenerates maximally, and leads to an $N=1$
vacuum on introducing an adjoint mass term. As we saw earlier this
singularity appears as a consequence of mutually local monopoles
becoming simultaneously light.
We have also seen, using
the effective Lagrangean at this point, that in the
presence of quark masses this special point gives rise to ${\cal N}_2$
ground states
belonging to the so-called second group of vacua with approximate
$\USp(2n_f)$ symmetry. Let us now see how this group of vacua shows up
in the language of the hyperelliptic curves after a mass perturbation.

Since the special point is at the root of an $r$-branch with $r=\tilde
n_c$, and a gauge symmetry enhanced to $SO(\tilde n_c)$, we must set $
r= n_f - {n_c\over 2} + 2 \,$ $\,\phi_a$'s to zero when $n_c$ is even
and $ r= n_f - {n_c -3 \over 2} \,$ $\,\phi_a$'s to zero when $n_c$ is
odd.

For even $n_c$ the curve at this point becomes
\eqn\specialcurve{ y^2 = x^{2 n_f-n_c +5} \, \left[ \prod_{a=1}^{
n_c-n_f-2} ( x-\phi_a^2)^2 - 4 \Lambda^{2(n_c-2-n_f) } x^{n_c-n_f-2}
\, \right].}
As argued in \arplsha\ the remaining $\phi_a$'s are given by $
\phi_a^2= (\omega, \omega^2, \ldots, \omega^{n_c-n_f-2}) \Lambda^2, $
where $\omega = \exp{ \pi i \over n_c-n_f-2} $.  With this choice of
the Coulomb branch VEVs, the expression in the brackets in
Eq.~\specialcurve\ becomes a perfect square,
\eqn\perfsq{ y^2 = x^{2 n_f-n_c +5} \, \left( x^{ n_c-n_f-2} - \Lambda^{2
(n_c-n_f-2) } \right)^{2},}
implying maximal degeneracy of the Riemann surface.  Similarly, for
odd $n_c$, we set $ r= n_f - {n_c -3 \over 2} \,$ $\,\phi_a$'s to
zero, and the curve becomes
\eqn\odd{ y^2 = x^{2 n_f-n_c +4} \, \left[ \prod_{a=1}^{ n_c-n_f-2} (
x-\phi_a^2)^2 - 4 \Lambda^{2(n_c-2-n_f) } x^{n_c-n_f-2} \, \right].  }
By choosing the remaining $\phi_a$'s as above, the curve takes the
form
\eqn\rem{ y^2 = x^{2 n_f-n_c +4} \, \left( x^{ n_c-n_f-2} - \Lambda^{2
(n_c-n_f-2) } \right)^{2}.}

Note that in both cases the order of the zero at $x=0$ is odd,
implying that the mass splitting of this point gives just one class of
vacua, as in the $USp(2n_c)$ case \ckm.  Therefore upon mass
perturbation, all large zeros remain doubled, near the roots of $x^{
n_c-n_f-2} - \Lambda^{2 (n_c-n_f-2) }=0$, with linear shifts which are
uniquely determined.

In order to find the number of vacua, it suffices to study the small
branch points, i.e. the perturbation of the roots at $x=0$ by the
quark masses.  Consequently we may ignore $x-$dependence of the
factors $ ( x-\phi_a^2)^2$ where the unperturbed $\phi_a$'s are large
(comparable to the nonperturbative scale).  Now the small $\phi$'s
must be such that
\eqn\smallphi{ y^2 = \Lambda^{4(n_c-2-n_f) } \, x
\,\prod_{a=1}^{{\tilde n}_c/2 } (x- \phi_a^2)^2 - 4
\Lambda^{2(n_c-2-n_f) }
x^{3} \prod_{i=1}^{n_f} (x-m_i^2), \quad n_c= {\hbox{\rm even}},} or
\eqn\smallphione{y^2 = \Lambda^{4(n_c-2-n_f) } \, x \,
\prod_{a=1}^{({\tilde n}_c-1)/2 } (x- \phi_a^2)^2 - 4
\Lambda^{2(n_c-2-n_f) } x^{2} \prod_{i=1}^{n_f} (x-m_i^2), \quad n_c=
{\hbox{\rm odd}},}
is maximally singular.  These are precisely the curves of the IR free
$SO({\tilde n}_c)$ theories with $n_f$ hypermultiplets having generic
nonzero masses.  Potential $N=1$ vacua can now be identified by
choosing $r$ of the $\phi_a^2$'s to match $r$ of the distinct quark
masses $m_i^2$.  This can of course be achieved in ${}_{n_f}C_r$ ways. 
Then, for $|x|\ll m_i^2,\;(i=1,\ldots r)$ the curve has the form
\eqn\evenpure{ y^2 = \Lambda^{4(n_c-2-n_f) } \, x
\,\prod_{a=1}^{({\tilde n}_c/2-r) } (x- \phi_a^2)^2 \prod_{i=1}^{r}
(m_i^4) - 4 \Lambda^{2(n_c-2-n_f) }
x^{3} \prod_{i=1}^{n_f} (m_i^2), \quad n_c= {\hbox{\rm even}},} when
there are an even number of colors.  Up to overall multiplicative
constants this is the curve for a pure $N=2$ gauge theory with
$SO(\tilde n_c-2r)$ gauge symmetry.  This theory will obviously yield
$w(n_c-2r)$ vacua upon deforming to $N=1$.  Analogous arguments apply
for the $n_c$ odd case and therefore we have shown that the special
point of APS does indeed give rise to
\eqn\absentbisbis{{\cal N}_2 = \sum_{r}^{[{\tilde n}_c/2] } w({\tilde
n_c}-2r) \, {}_{n_f}\!C_r + {}_{n_f}\!C_{{\tilde n}_c/2}}
$N=1$ ground states.

In all the cases analyzed above the location of the Chebyshev vacua
(e.g. number of vanishing $\phi_a$'s) is simply not compatible with
that of the special point.  For instance, for $n_f$ even, $n_c$ even,
the coincidence of the two points would imply $n_f - {n_c \o 2} + 2 =
{n_f \o 2} +1$, which is an impossible relation (since we limit to AF
cases $n_f < n_c -2$.)  This means that the Chebyshev points and the
special points are distinct and well-separated in the $N=2$ moduli
space.

\subsec{Summary}

In all cases we find that each of the $n_c-n_f-2$ Chebyshev points  on quantum moduli space gives rise, upon  generic $m_i$ perturbation, to $2^{n_f}$  singularities of the desired type, so that the total  number of $N=1$ vacua of the first group is
\eqn\groupone{{\cal N}_1 =(n_c-n_f -2) \,  2^{n_f}.}
The situation is rather similar to the  case of  $USp(2n_c)$ theories.
Around the special point (baryonic like root), there are ${\cal N}_2$ singularities ($N=1$ vacua).      ${\cal N}_1 + {\cal N}_2$ coincides with the total number of the  vacua found  from the semiclassical analysis and and at  large $\mu. $

%: Semiclassical monopole multiplets and singularities of QMS
\newsec{Semiclassical monopole multiplets and singularities of QMS}

In their seminal papers on the exact solution of $N=2$ supersymmetric gauge theories, Seiberg and Witten assumed that the monopoles which become massless at a singularity of the quantum moduli space are smoothly related to the massive semiclassical monopole of 't Hooft and Polyakov present at large $u= \bra \Tr \Phi^2 \ket$. 
Such a connection has been confirmed in $SU(2)$ theory with nonvanishing flavors, through the study of the flavor multiplet structure and fractional quark numbers of these monopoles \ktcp,\GKT.

The situation is however much less obvious in a more general class of $N=2$ gauge theories studied in \ckm\ and here.  
In generic $SU(n_c)$ and $USp(2n_c)$ theories, it was found that at the singularities of QMS which survive the $N=1$ perturbation $\mu \Tr \Phi^2 |_F$ the massless states are either dual quarks and flavor singlet monopoles or nonlocal set of dyons. 
At generic $r$- vacua (in the notation of \ckm) the flavor symmetry is broken by condensation of the dual quarks. Flavor multiplets of massless monopoles relate smoothly to semiclassical 't Hooft - Polyakov monopoles only in a restricted set of vacua (namely $r=1$, in the notation of \ckm). 
It was argued in \ckm ~ that the dual quarks appearing in the generic $r$- vacua of $SU(n_c)$ are the ``baryonic components'' of the semiclassical monopoles in the $r$- antisymmetric representation of the flavor $SU(n_f)$ group.

Both in $SU(n_c)$ and $\USp(2n_c)$ gauge theories,  an agreement was found between the number of  $N=1$ vacua in the first group and the total multiplicity of semi-classical monopole states, both of which turn out to be of order $2^{n_f}$ times a discrete symmetry factor which depends on the gauge group.

We have already seen that the first group of vacua of the $SO(n_c)$ theory with equal quark masses is in the same universality class as the $r$- vacua discussed above. Therefore, it would be interesting to understand the relationship between the light degrees of freedom (dual quarks) and the semiclassical monopole flavor multiplets of the $SO(n_c)$ gauge theory.
However in the generic $SO(n_c)$ theories studied in this paper, a puzzle emerges. There appears to be an order-of-magnitude discrepancy between the multiplicity of the semiclassical monopole states in the massless theory (which turns out to be $\sim 2^{2n_f}$, as shown below) and the number of the singularities corresponding to the $N=1$ vacua (which is $ \sim 2^{n_f}$).  Since the number of the latter is well-defined only for generic, nonvanishing bare quark masses, we are not facing any paradox here. 
It is nonetheless interesting to attempt to understand the origin of the difference between the situation in the $SO(n_c)$ theory on the one hand and $SU(n_c) $ or $\USp(2n_c)$ theories on the other.

The flavor contents of the semiclassical monopole states in the latter theories have been studied in an Appendix of \ckm.
For $SO(n_c)$ gauge group, we first observe that in the massless theory (i.e. with vanishing bare quark masses) each Dirac fermion has {\it two } zero modes in the background of the 't Hooft - Polyakov monopole, in contrast to the $SU(n_c)$ and $\USp(2n_c)$ theories.  
The doubling of the zero modes for each Dirac fermion follows from the symmetry of the classical equations.\foot{One of the authors (K.K.) thanks D. Kaplan for discussions on this point.} 
Namely, the Dirac equations for $2 n_f$ fermions $\psi_1, \ldots, \psi_{n_f}, {\tilde \psi}_1, \ldots, {\tilde \psi}_{n_f} $ are invariant under $\USp(2n_f)$ transformations generated by $\pmatrix {B & A \cr A^* & -B^T } $, where $B^{\dagger}= B; \,\, A^T=A$.  
Suppose that one zero mode for the Dirac pair $(\psi_1, {\tilde \psi}_1) = ( i \eta(r), \eta(r)) $ is found.  Now, $\USp(2n_f)$ transformations generated by $A_{11}, $ $A_{11}^{*},$ which act as an $SU(2) \subset \USp(2n_f)$, contain elements
\eqn\usp{ e^{i \alpha \tau_{1,2}/2} = \cos {\alpha \o 2} + i \,
\tau_{1,2} \, \sin {\alpha \o 2},}
which act on the $(\psi_1, {\tilde \psi}_1)$ subspace.  
By choosing $\alpha= \pi$, the above elements become proportional to
\eqn\uspelements{ \pmatrix{ 0 & 1 \cr 1 & 0}, \qquad {\hbox {\rm or}}
\qquad \pmatrix{ 0 & 1 \cr -1 & 0},}
showing that if $(\psi_1, {\tilde \psi}_1) = ( i \eta(r), \eta(r))$ is a solution, so is $ (\psi_1, {\tilde \psi}_1) = (- i \eta(r), \eta(r))$.
Of course, the transformations generated by $U(n_f) \subset \USp(2n_f)$ give rise to the zero modes for other flavors, in the standard way. 
In all, there are $2n_f$ zero modes.  
The doubling of the zero modes in the massless theory can be also established by the study of Callias' index theorem for $SO(n_c)$ theories, which we summarize in Appendix A. 
Quantization of the fermions introduces then $2n_f$ pairs of creation and annihilation operators in the zero mode sector.  
One thus expects to find $2^{2n_f}$ monopole states semiclassically. 
We now turn to the nontrivial task of explicitly constructing these monopole states in terms of the creation operators.

Call the $2n_f$ zero mode operators $b^i_{\alpha}$, where $i=1,2,\ldots, n_f,\,$ $\alpha=1,2$.  Let us define $2 n_f$ operators by
\eqn\defop{ c^{n_f + i}_{\alpha} = (b^i_{\beta})^{\dagger}
\epsilon^{\alpha \beta}, \qquad c^i_{\alpha}=b^i_{\alpha}, \qquad
(i=1,2,\ldots, n_f),}
where $ \epsilon^{21} =1.$ (Of course, only $2n_f$ operators among
$c^i_{\alpha}$ and their adjoints are independent.)  The standard
quantization conditions
\eqn\commut{ \{ b^i_{\alpha}, (b^j_{\beta})^{\dagger} \}= \delta^{ij}
\delta_{\alpha \beta},\qquad (i,j=1,2,\ldots, n_f)}
then translate into
\eqn\relations{\{c^i_{\alpha}, c^j_{\beta}\}= J^{ij} \epsilon^{\alpha
\beta}, \qquad (i=1,2,\ldots, 2 n_f),}
which is  indeed   invariant under  $\USp(2n_f) $.
In terms of  $ c^i_{\alpha},    c^{i \dagger}_{\alpha}$    the
$\USp(2n_f)$    charges       are
\eqn\uspcharges{\pmatrix{c^{1 \dagger} &  \ldots  &   c^{2n_f 
\dagger}}_{\alpha}
\pmatrix {B &   A \cr
A^* &    -B^T }     \pmatrix {c^{1 }  \cr \vdots \cr     c^{2n_f  
}}_{\alpha},}
where $\pmatrix {B & A \cr A^* &    -B^T } $ is the standard   $\USp(2n_f)$   generators with   $B^{\dagger}=  B; \,\, 
A^T=A$.   Equivalently,   in terms of the original
independent operators   they are:
\eqn\uspgene{\eqalign{  &  \pmatrix {b_1^{1 \dagger}  &   \ldots & 
b_1^{n_f \dagger} &
b_2^1 & \ldots  b_2^{n_f}}
\pmatrix {B &   A
\cr A^* &    -B^T }      \pmatrix {b_1^{1 } \cr  \vdots \cr b_1^{n_f
}\cr b_2^{\dagger 1} \cr \vdots \cr b_2^{\dagger n_f}}    \cr
&+
   \pmatrix {b_2^{1 \dagger}  &   \ldots & b_2^{n_f \dagger} & -b_1^1 
&
\ldots  -b_1^{n_f}}
\pmatrix {B &   A
\cr A^* &    -B^T }      \pmatrix {b_2^{1 } \cr  \vdots \cr b_2^{n_f
}\cr - b_1^{\dagger 1} \cr \vdots \cr - b_1^{\dagger n_f}}.
}}
It is convenient to introduce also ``$SU(2)$" generators, even though  they are not symmetry operators  of the full quantum theory:
\eqn\tauplus{  {\tau}_+= \sum_{i=1}^{ n_f}    b_1^{\dagger i } b_2^i 
=  {1 \o 2} \sum_{i=1}^{ 2 n_f}    c_1^{\dagger i } c_2^i =
\sum_{i=1}^{ n_f}    c_1^{\dagger i } c_1^{\dagger  n_f + i  } =  { 
1\o 2} J_{ij}  c_1^{ \dagger i } c_1^{\dagger j  };}
\eqn\tauminus{{\tau}_-=  \sum_{i=1}^{ n_f}    b_2^{\dagger i } b_1^i 
=  -  {1 \o 2} \sum_{i=1}^{ 2 n_f}    c_2^{\dagger i }
c_1^i  =
\sum_{i=1}^{ n_f}    c_1^{n_f+ i } c_1^{ i  } = -  { 1\o 2} J_{ij} 
c_1^{ i } c_1^{  j  } ;}
\eqn\tauthree{{\tau}_3=   {1 \o 2}    \sum_{i=1}^{n_f}  (b_1^{\dagger 
i}  b_1^i -  b_2^{\dagger i}   b_2^i ) =
  {1 \o 4}    \sum_{i=1}^{n_f}  (c_1^{\dagger  i}  c_1^i - 
c_2^{\dagger i}   c_2^i ) ={ 1\o 2} J_{ij}  c_1^{ i } c_2^{  j  },
}
for the purpose of constructing the semiclassical spectrum of 
monopoles.
Note that  all of $\tau^i$   are   singlets of $\USp(2n_f)$.  
Vice versa,  the generators of $\USp(2n_f)$ in \uspgene, are obviously  all $SU(2)$ singlets,  showing that   $\USp(2n_f)$ and  $SU(2)$ commute with each other.

One can construct the Fock space of states by treating any set of  $2n_f$ independent operators
as annihilation operators and defining the vacuum with respect to them.
In order to  see  the  multiplet structure of $\USp(2n_f)$  (which is the true symmetry of the system)   it is convenient to introduce  a ``vacuum" state $|0\ket$ defined by
\eqn\vacuum{c_{1}^i |0\ket=0, \qquad \quad i=1,2,\ldots 2 n_f.
}
Various $\USp(2n_f)$ tensors can then be constructed as follows:

\item{i)}  $USp(2n_f)$ singlets,
\eqn\uspsing{|0\ket, \,\, \tau_+ |0\ket,\,\, \tau_+^2 |0\ket,\,\, 
\ldots, \tau_+^{n_f} |0\ket:}
they form a ``spin" ${n_f \o 2}$  multiplet, with multiplicity $n_f+1.$
Note that $\tau_+^{n_f +1} |0\ket =0$;
\item{ii)}  $ {\underline {2n_f}} $ of    $USp(2n_f)$,
\eqn\forinst{c_1^{\dagger i}  |0\ket, \,\, \tau_+   c_1^{\dagger i} 
|0\ket,\,\, \tau_+^2   c_1^{\dagger i}  |0\ket,\,\, \ldots,
\tau_+^{n_f-1}   c_1^{\dagger i}  |0\ket:}
which form the ``spin"  ${n_f -1 \o 2}$ $SU(2)$ multiplet, with multiplicity ${}_{2n_f}C_{2} \times  n_f$;
\item{iii)}  Second rank antisymmetric irreps of $USp(2n_f)$, constructed from $ c_1^{\dagger i} c_1^{\dagger j}  |0\ket,$ by subtracting the singlet,
\eqn\secran{\left[ c_1^{\dagger i}  c_1^{\dagger j} -  {J_{ij} \o  2 n_f}  \left(\sum_{k\ell}  J_{k \ell}  \, c_1^{\dagger k} c_1^{\dagger \ell} \right)  \right]|0\ket,}
and  those obtained   by  acting on these states with $\tau_+$  up to  $n_f -2$ times.
The multiplicity is
\eqn\multip{( {}_{2n_f}C_{2}-{}_{2n_f}C_{0} ) \cdot (n_f-1);}
\item{iv)}   General rank $r$ antisymmetric irrep of $USp(2n_f)$, constructed from
\eqn\genrank{c_1^{\dagger i_1} c_1^{\dagger i_2}  \ldots  c_1^{\dagger i_r}  |0\ket,}
by subtracting all possible contractions with $J_{ij}.$
Other  rank $r$ antisymmetric irreps can be obtained by acting on these with $\tau_+$ up to  $n_f - r$ times.
For instance,   for $r=4$  the resulting multiplicity  of an irrep of  $\USp(2n_f)$  is
\eqn\forinst{ {}_{2n_f}C_{4}-  (  {}_{2n_f}C_{2} - {}_{2n_f}C_{0})  -  {}_{2n_f}C_{0} =    {}_{2n_f}C_{4}-    {}_{2n_f}C_{2}.}
In general, for  general $r$,  the multiplicity is
\eqn\kmultip{( {}_{2n_f}C_{r}-    {}_{2n_f}C_{r-2} )  \cdot (n_f - r +1),}
where the second factor  is due to the  ``$SU(2)$ spin".
\item{v)}  Finally,   the $n_f$-antisymmetric tensor,
\eqn\muraset{c_1^{\dagger i_1}  c_1^{\dagger i_2}  \ldots  c_1^{\dagger i_{n_f} }  |0\ket - {\rm contractions}}
with multiplicity,
\eqn\spinzerom{{}_{2n_f}C_{n_f}-    {}_{2n_f}C_{n_f-2}.}

\noindent Clearly, the maximum rank of the antisymmetric $USp(2n_f)$  tensor  constructed  this way is $n_f$,  so the total multiplicity of the states is
\eqn\totalmu{\eqalign{& \sum_{r=2}^{n_f} 
({}_{2n_f}C_{r}-{}_{2n_f}C_{r-2}) (n_f-r+1)
   + {}_{2n_f}C_{1} \cdot  n_f + {}_{2n_f}C_{0} \cdot   (n_f+ 1)  \cr
=& \sum_{r=0}^{n_f} {}_{2n_f}C_{r} (n_f-r+1)
- \sum_{r=0}^{n_f-2} {}_{2n_f}C_{r} (n_f-r-1) \cr
=& 2 \sum_{r=0}^{n_f-2} {}_{2n_f}C_{r}
+2  \cdot    {}_{2n_f}C_{n_f-1} + {}_{2n_f}C_{n_f}  = 
\sum_{r=0}^{2n_f} {}_{2n_f} C_{r}  =     2^{2n_f},}}
as expected.

Semiclassically, these monopole are all massive, and  only the degeneracy within  the same  $\USp(2n_f)$ multiplet is expected to survive the full quantum effects.
When bare quark masses are added to the theory, each  $\USp(2n_f)$ multiplet above  decomposes into a sum of  $SU(n_f)$ multiplets.

The number of the singularities  corresponding to $N=1$ vacua we found    ($\sim 2^{n_f}$)      is much  smaller  than that of the multiplicity of the semiclassical  monopole states,   of the order of 
$2^{2n_f}$, in contrast to the situation in $SU(n_c)$ and $\USp(2n_c)$   theories.
This fact can be ``understood''  ~   from  the details of our analysis in section (5.1) on the mass
perturbation.  
Indeed, one component of  the adjoint VEVS   $\, \phi_a$ was   found to  vanish in all cases  for the nonvanishing quark masses, see  \pert, \vanish, \vanishing ~and  \kieru.
According to the Callias index theorem (Appendix A) it means that there is only one zero mode for each Dirac fermion  in such a background. 
The surviving zero mode is perfectly normalizable due to the
exponential damping factor at the spatial infinity, while the other 
would-be zero mode becomes non-normalizable for the case
of a finite mass.  (The massless limit is tricky as there is no exponential
damping factor with $\phi=0$ and the would-be zero mode
is  marginally non-normalizable.)
Under these circumstances, one can construct only monopole states in the representations of $U(n_f) \subset \USp(2n_f)$  group, not of the full group $\USp(2n_f)$, as can be seen from \uspgene -- $A_{ij}$   being symmetric,  one needs two independent zero modes $b_{\alpha}^j, \, (\alpha=1,2)$ to get  nonvanishing Noether charges of  $\USp(2n_f)/U(n_f)$ in the zero mode sector).
This precisely corresponds to what the authors of \eguhorione~found, namely that in the massive $SO(n_c)$ theory the  SCFT's of the first group are in the same universality class as those found in the $SU(n_c)$  gauge theory.
Different $r$ vacua are described by an effective $ SU(r) \times U(1)^{[{n_c \over 2}]-r+1}$ theory with $n_f$ dual quarks in the fundamental representation of $SU(r)$. (See Table 2.) The sum of these $N=1$ vacua then gives (\equalnone) the total number of singularities $(n_c- n_f -2) \, 2^{n_f}.$

The above discussion demonstrates that the counting of monopole
multiplicity is a subtle issue in the massless quark limit, and clarifies to some extent the differences between the $SO(n_c) $ and $SU(n_c)/ \USp(2n_c)$ theories with regard to the flavor structure of semiclassical monopole  states. Unfortunately it does not shed   much light  on the dynamical  details of these vacua.    
The physics of these ``deformed SCFT'' vacua, the precise mechanism of symmetry breaking  and confinement  and the role of the magnetic monopoles therein remain to be further  elucidated.

\nref\callias{C. Callias, ``Axial Anomalies and Index Theorems on Open
Spaces," \cmp{62}{1978}{213}; E. Weinberg, ``Fundamental Monopoles 
in Theories with Arbitrary Symmetry Breaking," \np{B203}{1982}{445}; J. de Boer, K. Hori and Y. Oz,
``Dynamics of N=2 Supersymmetric Gauge Theories in Three Dimensions,"
hep-th/9703100, \np{B500}{1997}{163}.}

%: Acknowledgments
\newsec{Acknowledgments}

Two of the authors (K.K. and S.P.K.) would like to thank the Particle Theory Group at the University of Washington where part of this work was done. 
G.C. and S.P.K. would like to thank Nick Dorey for useful discussions.
S.P.K. was partially supported by DOE grant DE FG03-96ER40956 and PPARC grant PPA/G/0/1998/00611. G.C. was supported by PPARC grant PPA/G/8/1998/00511. 
H.M. was supported in part by the Director, Office of Science, Office 
of High Energy and Nuclear Physics, Division of High Energy Physics of
the U.S. Department of Energy under Contract DE-AC03-76SF00098 and
in part by the National Science Foundation under grant PHY-95-14797.

\listrefs

\appendix {A}{Callias Index Theorem for $SO(n_c)$ Theories}

Start from the index theorem for the Dirac zero modes \callias
\eqn\callias{{\bf N} = \sum_{a=1}^{\ell} n_a \sum_{\bf w} {{\bf w}
\cdot \beta^{(a)} \o \beta^2} {\hbox {\rm sgn}} [({\bf w} \cdot \phi)
- m],}
where $\ell$ and $\beta^{(a)}$ are the rank and simple roots of the
gauge group considered, ${\bf w}$ are weight vectors of a given
representation to which the fermion belongs, and $n_a$ is the integer
monopole magnetic charge associated to the $a$-th $U(1)$ factor of the
Cartan subgroup.
For $SO(2 \ell )$, the simple roots are ${\bf e}_i-{\bf e}_{i+1}$
($i=1,2,\ldots, \ell-1$) and ${\bf e}_{\ell-1}+{\bf e}_{\ell}$, where
${\bf e}_i$'s are orthonormal vectors in an $\ell$ dimensional
Euclidian space, and one finds
\eqn\soncallias{ \eqalign {    {\bf    N} = &   \sum_{a=1}^{\ell} 
n_a  \sum_{\bf w} {{\bf w} \cdot \beta^{(a)}  \o  \beta^2}
{\hbox {\rm sgn}}      [({\bf w} \cdot \phi) - m]     \cr
=& { 1\o 2 } \, [ \,  n_1  \,\{{\hbox {\rm sgn}} (\phi_1-m) - {\hbox 
{\rm sgn}}
  (-\phi_1-m) -{\hbox {\rm sgn}} (\phi_2-m)+ {\hbox
{\rm sgn}} (- \phi_2-m)  \}   \cr
+ &  n_2  \{{\hbox {\rm sgn}} (\phi_2-m) -{\hbox {\rm sgn}} 
(-\phi_2-m) - {\hbox {\rm sgn}} (\phi_3-m)+ {\hbox {\rm sgn}}
(- \phi_3-m)  \} \cr
+ &    \ldots  \quad \ldots  \quad  \ldots   \cr
+ &  n_{\ell-1}    \{{\hbox {\rm sgn}} (\phi_{\ell-1}-m) - {\hbox 
{\rm sgn}}
(-\phi_{\ell-1}-m) - {\hbox {\rm sgn}} (\phi_{\ell}-m) +
{\hbox {\rm sgn}} (- \phi_{\ell}-m)  \}  \cr
+ &  n_{\ell } \, \{{\hbox {\rm sgn}} (\phi_{\ell-1}-m) - {\hbox {\rm 
sgn}} (-\phi_{\ell-1}-m) +  {\hbox {\rm sgn}} (\phi_{\ell}-m) -
{\hbox {\rm sgn}} (- \phi_{\ell}-m)  \} \, ].}} 
The reason for the dependence on signs of $(\pm \phi - m)$ is
because the would-be zero mode behaves as $e^{-(\pm \phi-m) r}$ at
the spatial infinity and the zero mode exists only if it is normalizable.
For example,  for  $SO(4)$,    $\ell=2,$  the index is  given by the formula
\eqn\haicap{\eqalign {    {\bf    N}    =& { 1\o 2 } \, [ \,  n_1 
\,\{{\hbox {\rm sgn}} (\phi_1-m) - {\hbox {\rm sgn}} (-\phi_1-m) -
{\hbox {\rm sgn}} (\phi_2-m)+ {\hbox {\rm sgn}} (- \phi_2-m)  \} \cr
+&  n_{2} \, \{ {\hbox {\rm sgn}} (\phi_{1}-m) - {\hbox {\rm sgn}} 
(-\phi_{1}-m) +  {\hbox {\rm sgn}} (\phi_{2}-m) -
{\hbox {\rm sgn}} (- \phi_{2}-m)  \} \, ],}}
so
\eqn\region{\phi_1 >  \phi_2  > m  > 0  \quad \to \quad   {\bf    N}= 
2  \,  n_{2};}
\eqn\regio{\phi_1   > m >  \phi_{2}  > 0  \quad \to \quad   {\bf 
N}=  \, n_{1} + n_2;}
\eqn\reg{m >   \phi_1 > \phi_2  > 0    \quad \to \quad   {\bf    N}=  
0.}
For general $SO(2\ell),$   the number of the zero modes depends on 
the field configuration as:
\eqn\genreg{\phi_1> \phi_2 > \ldots > \phi_{\ell} > m  > 0  \quad \to 
\quad   {\bf    N}= 2 \,  n_{\ell};}
\eqn\genrgio{\phi_1> \phi_2 > \ldots > \phi_{\ell-1} >m > 
\phi_{\ell}  > 0  \quad \to \quad   {\bf    N}=  n_{\ell-1} + 
n_{\ell};}
\eqn\genren{\phi_1> \phi_2 > \ldots > \phi_{\ell-2} >m > 
\phi_{\ell-1}  > 0  \quad \to \quad   {\bf    N}=  n_{\ell-2};}
\eqn\genrens{\ldots \quad  \ldots \quad  \ldots  }
\eqn\genma{\phi_1>  m > \phi_2 > \ldots >  \phi_{\ell}  > 0  \quad 
\to \quad   {\bf    N}=  n_{1};}
\eqn\geninso{m>   \phi_1> \phi_2 > \ldots >  \phi_{\ell}  > 0  \quad 
\to \quad   {\bf    N}=  0.}

For $SO(2 \ell +1 )$,
the simple roots are  ${\bf e}_i-{\bf e}_{i+1}$
   ($i=1,2,\ldots, \ell-1$)  and ${\bf e}_{\ell}$: the index formula 
is
\eqn\sonplus{\eqalign{  {\bf N}
=&  { 1\o 2 } \, [ \,  n_1  \,\{ {\hbox {\rm sgn}} (\phi_1-m) - 
{\hbox {\rm sgn}} (-\phi_1-m) - {\hbox {\rm sgn}} (\phi_2-m)+ {\hbox
{\rm sgn}} (- \phi_2-m)  \cr
  +&    \ldots  \quad  \ldots  \quad  \ldots    \cr
  +&  n_{\ell-1}    \{ {\hbox {\rm sgn}} (\phi_{\ell-1}-m) - {\hbox 
{\rm sgn}} (-\phi_{\ell-1}-m) -{\hbox {\rm sgn}} (\phi_{\ell}-m) +
{\hbox {\rm sgn}}    (- \phi_{\ell}-m)  \} \cr
+& 2  n_{\ell } \, \{ {\hbox {\rm sgn}} (\phi_{\ell}-m) -
{\hbox {\rm sgn}} (- \phi_{\ell}-m)  \} \, ].}}
In this case    the zero mode multiplicity depends on the 
configuration and on the monopole charges as:
\eqn\regsonpo{\phi_1> \phi_2 > \ldots > \phi_{\ell} > m  > 0  \quad 
\to \quad   {\bf    N}= 2 \,  n_{\ell};}
\eqn\regs{     \phi_1> \phi_2 > \ldots > \phi_{\ell-1} >m > 
\phi_{\ell}  > 0  \quad \to \quad   {\bf    N}=  n_{\ell-1};}
\eqn\regso{     \phi_1> \phi_2 > \ldots > \phi_{\ell-2} >m > 
\phi_{\ell-1}  > 0  \quad \to \quad   {\bf    N}=  n_{\ell-2};}
\eqn \regson{ \ldots \quad  \ldots \quad  \ldots }
\eqn\regsonp{     \phi_1>  m > \phi_2 > \ldots >  \phi_{\ell}  > 0 
\quad \to \quad   {\bf    N}=  n_{1};}
\eqn\regnn{   m>   \phi_1> \phi_2 > \ldots >  \phi_{\ell}  > 0  \quad 
\to \quad   {\bf    N}=  0.}
We note that  both in  $SO(2\ell)$ and   $SO(2\ell+1)$ theories,  the 
doubling of the zero modes of a given Dirac fermion
requires
that all the $\phi$'s to be larger than  the mass $m$.
Therefore the massless limit and the monopole spectrum discussed
in Section 6 is valid when all $\phi$ are non-vanishing, but
when one of the $\phi$ vanishes the would-be zero mode loses the exponential
damping factor at the spatial infinity and 
the number of zero modes is ill-defined.  It is not clear what the spectrum
of semi-classical monopole is when one (or more) of $\phi$ vanishes
for the massless quark case.

\end